\documentclass[traditabstract]{aa} % for the abstract without structuration 

\usepackage{graphicx}
\usepackage{txfonts}
\usepackage{natbib}
\usepackage{stfloats} % fix positioning of 2 column figures
\usepackage{fixltx2e} % fix bad ordering of 2 column figures

\usepackage{color}

\bibpunct{(}{)}{;}{a}{}{,} % to follow the A&A style
\newcommand{\Atwo}{\object{P/2010 A2}}

\newcommand{\newtext}[1] {{#1}}

\begin{document}

   \title{
P/2010~A2 LINEAR
}
 \subtitle{II: Dynamical dust modelling}

   \author{
     J.~Kleyna     \inst{1} 
\and %Jan models, data      %% kleyna@ifa.hawaii.edu
     O.~R.~Hainaut \inst{2} \and %oli                     %% ohainaut@eso.org
     K.~J.~Meech\inst{1} %%\and %Karen  co               %% meech@ifa.hawaii.edu
%%     X.~Xxxx\inst{3}
%% Jan - you are in charge! add/remove as needed.
%
%
%
%%     A.~Zenn       \inst{1}      %Tonny  models           %% anthonyz@hawaii.edu
%%     G.~Sarid      \inst{1} \and %Gal   model             %% galahead@ifa.hawaii.edu
%%     B.~Hermalyn   \inst{3} \and %% Brendan Hermalyn Brown University
%%     H.~Hsieh      \inst{1} \and %Henri obs               %% hsieh@ifa.hawaii.edu
%%     J.~Licandro   \inst{4} \and %Javier ? do we use his data %%% jlicandr@iac.es
%%     H.~M.~Kaluna  \inst{1} \and %Heather  Observer       %% kaluna@ifa.hawaii.edu
%%     P.~Schulz     \inst{3} \and %Peter Schulz      %% peter_schultz@brown.edu
%%     G.~Tozzi      \inst{5} \and %GianPaolo proposal      %% tozzi@arcetri.astro.it
%%     G.~Trancho    \inst{6} \and %Gelys Gemini support    %% gtrancho@gemini.edu
%%     J.~Pittichov\'a \inst{1} \and %Jana                    %% jana@IfA.Hawaii.Edu
%%     B.~Yang       \inst{1}        %Bin                     %% yangbin@ifa.hawaii.edu
                                                          %% caterpillar.yang@gmail.com 
          }

   \institute{
     %1
     Institute for Astronomy (IfA), University of Hawai`i, 2680 Woodlawn Drive, Honolulu, HI 96822, USA\\
\email{kleyna@ifa.hawaii.edu} 
     \and
     %2
     European Southern Observatory (ESO), Karl Schwarzschild Stra\ss
     e, 85748 Garching bei M\"unchen, Germany   \email{ohainaut@eso.org}
     % \and
     % %3
     % Geological Sciences-Brown University, 324 Brook Street,
     % Providence RI 02912, USA     
     % \and 
     % %4
     % Instituto de Astrof\'{\i}sica de Canarias, c/V\'{\i}a Lactea
     %  s/n, 38200 La Laguna, Tenerife, Spain, and Departamento de
     %  Astrof\'{\i}sica, Universidad de La Laguna, E-38205 La Laguna,
     %  Tenerife, Spain 
     % \and 
     % %5
     %  INAF - Osservatorio Astrofisico di Arcetri, Largo E. Fermi 5,
     %  Firenze 50125, Italy 
     % \and 
     % %6
     % Gemini Observatory, Colina El Pino s/n, La Serena, Chile
        }

   \date{Accepted, {\it Astronomy and Astrophysics}}

   \abstract { \Atwo\ is an object on an asteroidal orbit that was
     observed to have an extended tail or debris trail in January
     2010. In this work, we fit the outburst of \Atwo\ with a conical
     burst model, and verify previous suspicions that this was a
     one--time collisional event rather than an sustained cometary
     outburst, implying that \Atwo\ is not a new Main Belt Comet
     driven by ice sublimation.  We find that the best--fit cone
     opening angle is $\sim 40^\circ$ to $\sim 50^\circ$ , in
     agreement with numerical and laboratory simulations of cratering
     events.  Mapping debris orbits to sky positions suggests that the
     distinctive arc features in the debris correspond to the same
     debris cone inferred from the extended dust.  From the velocity
     of the debris, and from the presence of a velocity maximum at
     around $15\rm\, cm\,s^{-1}$, we infer that the surface of A2
     probably has a very low strength ($\lesssim 1\,\rm kPa$),
     comparable to lunar regolith.

   }

  \maketitle

%
%________________________________________________________________
%________________________________________________________________
\section{Introduction}
%________________________________________________________________
%________________________________________________________________

\Atwo\ was discovered by LINEAR \citep{CBET2114} on
7~Jan~2010 on an orbit typical of a Main Belt asteroid, with a
Tisserand parameter $T_J = 3.6$ and orbital elements suggesting it
belongs to the Flora collisional family.  At the time of discovery, it
appeared as ``a headless comet with a straight tail, and no central
condensation'' \citep{IAUC9105}. A few days later, observers reported
an asteroid-like body connected to the tail by a narrow light bridge
\citep{IAUC9109}. \cite{2010IAUC.9109....3J} and 
\cite{2010CBET.2134....3L} interpreted the detached nucleus as the consequence
of an impact.

%------------------------------------------------------------------------------
\begin{figure*}
\includegraphics[width=18cm]{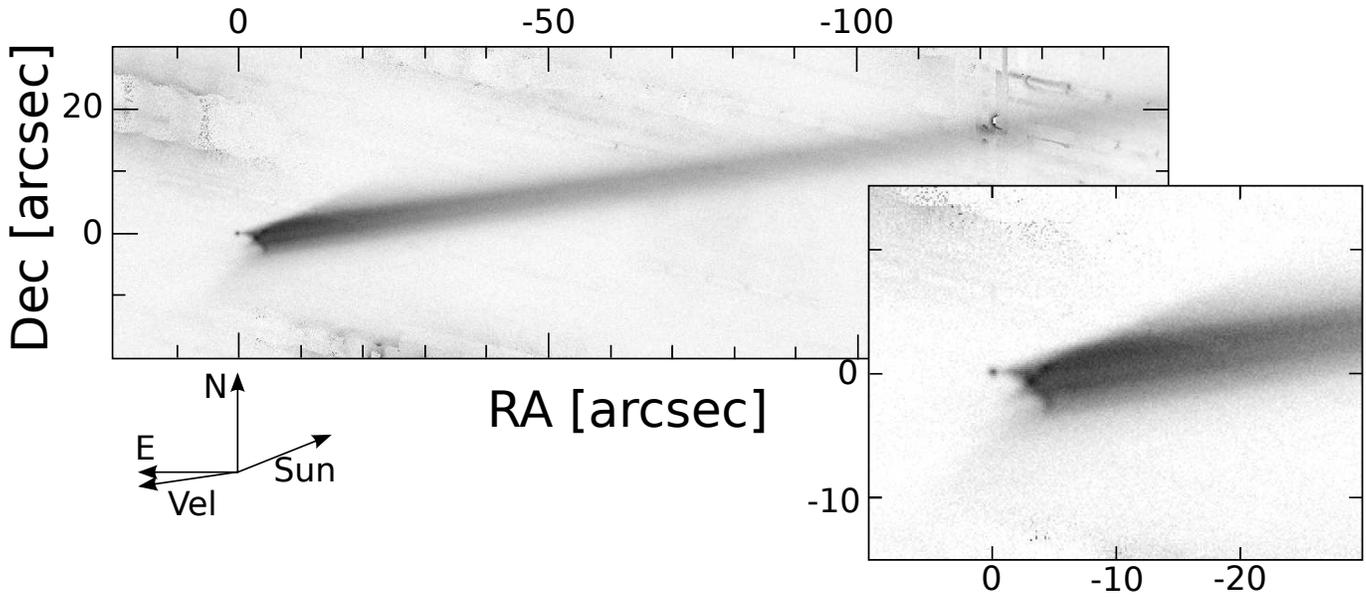}
\caption{\newtext{\Atwo, images from UT~19.5~Jan~2010 using Gemini North.
  The linear gray scale covers the range of (0--3) $\times 10^{-8} Af$
  (a dust proxy; see Paper I, Section 3.3.3). The positions of North (N)
  and East (E) are indicated, as are the anti-solar direction and the
  heliocentric velocity vector. From Paper I, Figure 2.e.}}
     \label{fig:image}%
\end{figure*}
%------------------------------------------------------------------------------

\citet{Mor+10} observed the comet with the \newtext{ Gran Telescopio Canarias, the
William Herschel Telescope,  and the Nordic Optical Telescope on La
Palma.} They modeled the observed tail using  water-driven cometary
activity extending over a period of several months. 

\citet{Jew+10} acquired a series of observations with the Hubble Space
Telescope over a long period from January to May 2010. The nucleus
appears not to be immediately surrounded by dust, which is further
confirmed by a fairly constant absolute magnitude over the span of
their observations. The detached tail is a narrow trail, striped with
very narrow, parallel striae that emanate from two sharp arcs
crossing to form a X.  From the evolution of the tail geometry, in
particular its orientation, they concluded that the dust release
occurred during a very short event that took place in
Feb.--Mar. 2009. They suggest that this event was caused either by a
collision or a spin up of the nucleus. 

\cite{Sno+10} acquired images of \Atwo\ using the camera onboard
the Rosetta space probe. The very different ---and favourable--
viewing geometry allowed them to constrain accurately the dust release
period to a very short burst around 10~Feb~2009.

We acquired a series of ground-based images using Gemini North and the
University of Hawai`i 2.2-m telescope on Mauna Kea, and the ESO New
Technology Telescope on La Silla. The observations, data processing
and overall analysis are presented in \citet[hereafter Paper
  I]{Hai+11}. Figure~\ref{fig:image} shows the deepest image of the
series.  The Gemini observations, which combined a very sharp image
quality with very deep surface brightness sensitivity, further confirm
that the nucleus is devoid of nearby dust. Assuming an albedo of
$p=0.11$ and a solar phase correction $G=0.15$ (values typical for
S-type asteroids, the most frequent members of the Flora family), its
magnitude converts to a radius $r=80$--$90\,\rm m$, in agreement with the
values reported by \citet{Mor+10}, \citet{Jew+10} and \citet{Sno+10}.

The analysis included thermal modelling of the nucleus,
which ruled out the presence of any water ice (as well as all more
volatile species) down to the center of the object, provided it
remained on its current orbit for more than a few million years. As
there is no reason to suspect that \Atwo\ has been recently
injected into its orbit, this excludes cometary activity as the
source of the dust release. 

\begin{figure}
\includegraphics[width=8cm]{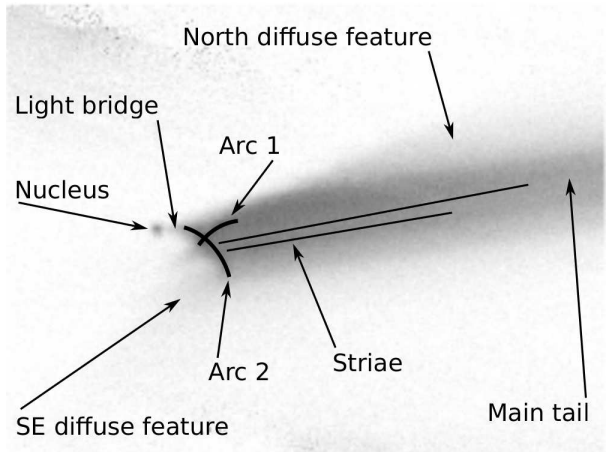}
\caption{Schematic of the main features of \Atwo, from Paper I.}
\label{fig:schematic}
\end{figure}

The data presented in Paper~I also show the features observed by
\citet{Jew+10}; they are summarized in Fig.~\ref{fig:schematic}.  The
tail-like dust trail was analyzed using the \citet{FP68} method,
constraining the dust release to a short period of time about one year
before the observations. This modelling, combined with direct
measurements of the tail, indicate that the dust grains had radii $a$
in the mm to cm range, with a size distribution in $a^{\sim -3.5}$,
and that the tail contained over $8\times 10^8$~kg of dust, assuming a
density $\rho=3\,000$~kg~m$^{-3}$ (typical value for a S-type
asteroid).

More advanced dust dynamical models were developed to investigate the
origin of the dust emission. While Paper I summarized the main
features of these models, this paper describes them in detail.
Sections \ref{sec:dust} and \ref{sec:dustcomplex} describes the models
used to study the overall envelope of the dust trail, while Section
\ref{sec:arcs} is devoted to the analysis of the X-shape arcs, labeled
Arc 1 and 2 in Fig.~\ref{fig:schematic}.

%------------------------------------------------------------------------------
%------------------------------------------------------------------------------

\section{Modeling the dust distribution - a simple binary envelope fit}
\label{sec:dust}

Our principal aim is to determine whether a simple single--burst model
can account for the most salient features of the \Atwo\ outburst, and
whether a burst corresponding to a collision event fits the data.  A
general single--burst model consists of an arbitrary differential
distribution of dust particles of solar radiation pressure coefficient
$\beta$ \footnote{$\beta$ is the ratio of radiation pressure to gravity, as
  defined in \cite{Hai+11}.} and velocity vector ${\bf v}_{\rm d}$,
given by $f({\bf v}_{\rm d},\beta)\,d{\bf v}\,d\beta$.  

Various approaches have been used for this type of general dust and
debris fitting.  For example, \cite{Jorda2007Icarus} modeled the Deep
Impact ejecta cone using a $\chi^2$ fit of a set of synthetic image
components, and obtained the dust size and velocity distributions.
Deep Impact had the advantage of a known ejecta geometry; if this
procedure were repeated with a completely unknown geometry, the
problem would become intractable using this image superposition
method, because the number of input images would be multiplied by the
number of possible geometries.  Moreover, the relatively clean and low
noise dataset of Deep Impact permitted the use of a quadratic
(Gaussian) fit, which is amenable to linear solutions.

\cite{Moreno2009ApJS} modeled comet 29P/Schwassmann--Wachmann using a
similar linear approach, finding that a simplified set of emission
regions combined with rotation gave a better fit than fixed sunward
emission.

We elect not to use these superposition methods for several reasons.
Because we wish to reconstruct the emission direction, we cannot
assume a direction like \cite{Jorda2007Icarus}.  A completely general
superposition solution permitting emission in cones spaced at
$10^\circ$ intervals in latitude and longitude, with 10 opening
angles, 10 velocities, and 10 dust sizes would require solving for
over half a million linear coefficients, an intractable computation
because linear problems scale with the cube of the number of
components.  Even if this were computable, it would be necessary to
impose normalization (smoothness) conditions that would dominate any
solution.

\begin{figure*}
\includegraphics[width=7cm, angle=270]{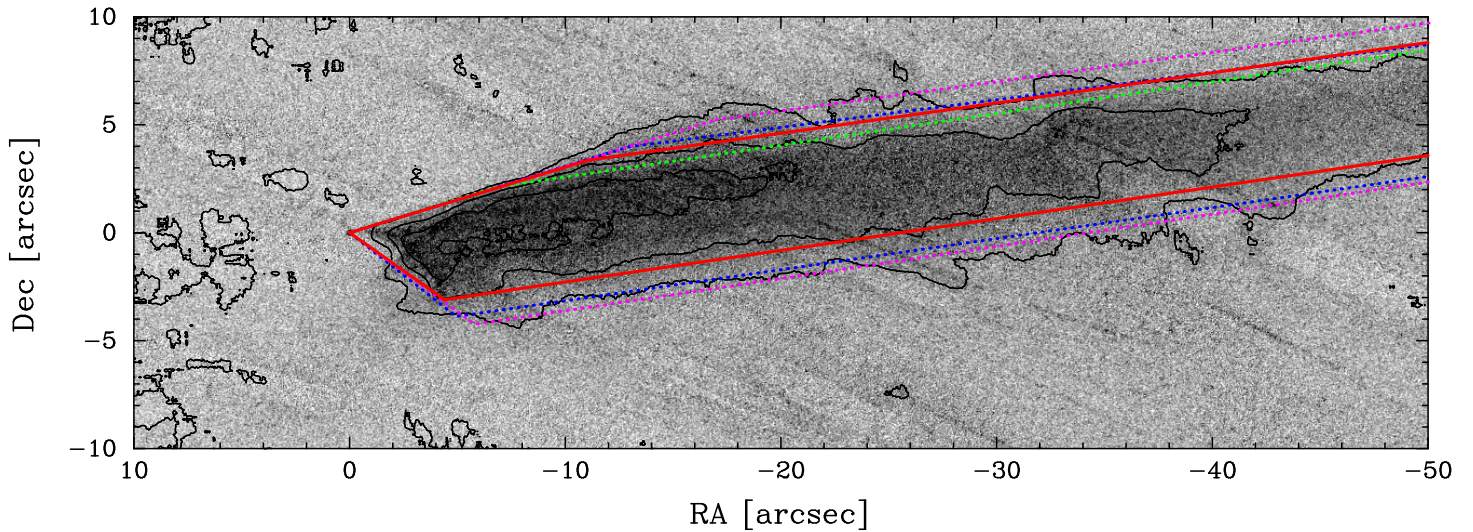}
\caption{Our best estimate (solid red line) of the dust envelope
  superimposed on the 29~Jan~2010 HST image, along with three other
  plausible envelopes (dotted lines in blue, green, and magenta).  The
  isophotal contours (black) are in rough agreement with the
  envelopes. The envelopes fit equally well on our Gemini image
  (Fig.~\ref{fig:image}).}
\label{fig:conemodelEnvelope}
\end{figure*}

Thus, instead of undertaking a general fit, we elect to fit a much
less complicated parametric model.  Our first simplification is to
assume that the outburst is a single hollow cone, with an opening
half--angle $\alpha_c$; for example, $\alpha_c=0$ corresponds to a
thin pencil--beam of debris.  \newtext {Such cones are typical of
  impact ejecta  (see
\cite{Richardson.Icarus.2007.ejecta}, hence R2007, and
\cite{HolsappleImactScalingReview1993} for overviews).}
 We define a right--handed orthogonal
coordinate system $(x_{A2},y_{A2},z_{A2})$ such that the nucleus is at
the origin, $x_{A2}$ points away from the Sun, $y_{A2}$ is in the
orbital plane and points in the sense of the orbital motion, and
$z_{A2}$ points out of the orbital plane.  There are corresponding
polar coordinates $\phi_{A2}$, the latitude angle from the orbital
plane toward $z_{A2}$, and $\theta_{A2}$, the angle in the
$x_{A2},y_{A2}$ plane starting in the $+x_{A2}$ direction.  For
example, the direction $\phi_{A2}=90^\circ$ points at the $+z_{A2}$
pole, and $\theta_{A2}=90^\circ, \phi_{A2}=0^\circ$ points in the
$+y_{A2}$ direction.

Our second simplification is to disregard completely the intensity of
the dust distribution, and consider only whether it fills the apparent
dust envelope of the observations (i.e., a {\sl binary} fit).  This means that we may ignore the
distributions of the dust size and velocity, and consider only their
minimum and maximum values.  By setting the minimum velocity and dust
size to be zero, assuming that dust size is independent of velocity,
and assuming that the minimum dust size escapes beyond the fitting
region, these four parameters reduce to the single parameter of
maximum dust velocity.  Hence the fitting procedure is reduced to
four dimensions: $\alpha_c, \phi_{A2}, \theta_{A2}, {v_{\rm
    d}}_{\rm max}$.

% We note that the velocity ${v_{\rm d}}_{\rm max}$ we fit is the
% velocity at a large distance from the nucleus, where the particle has
% already lost part of its initial velocity as a result of the weak
% attraction of the body.    The escape velocity was estimated to
% be $v_e\approx 10\,{\rm cm\,s^{-1}}$ \citep{Hai+11}, which is smaller
% than the velocities we deal with, and much smaller in energy
% terms than the $30\,{\rm cm\,s^{-1}}$ high end of our velocity range, 
% because $v_{\rm final}= \left({v_{\rm ejection}}^2 - {v_e}^2\right)^{1/2}$.

Our code contains two orbit integrators: a conventional Runge-Kutta
stepper, which  takes the gravity of the nucleus into account but
is slow; and an exact Keplerian solver that integrates the orbit in a
single step, but sees only the Sun's gravity and solar pressure.
Tests revealed that adjusting the ejection velocity by an assumed escape
velocity $v_e\sim 10\,{\rm cm\,s^{-1}}$ \citep{Hai+11}, according
to $v_{\rm final}= \left({v_{\rm ejection}}^2 - {v_e}^2\right)^{1/2}$
brought the Keplerian variant into very close observational agreement
with the full Runge-Kutta integrator, so the faster Keplerian code was
used throughout, even when taking into account the mass of the nucleus.

The fitting procedure is then to assume the outburst date of
\cite{Sno+10}, 10~Feb~2009 ($\pm 5$ days).  We
create a conical burst with a particular set of parameters $\alpha_c,
\phi_{A2}, \theta_{A2}, {v_{\rm d}}_{\rm max}$, integrate it to
the date of the observations, create a model image, and evaluate the
goodness of fit to the actual image. The penalty function of the fit
consists of $+1$ for each pixel of the model image that falls outside
the observed dust envelope, and $+1$ for each pixel inside the
observed envelope that is dust-free.  For pixels that fall within
$20\arcsec$ of the nucleus, we multiply the penalty by 10, to
enforce a good fit of the corner features at the cost of being lax
with the distant tail.  If the model dust perfectly fills the
envelope, and no dust falls outside, the penalty is zero.

The observed envelope is determined by finding, by eye, the limits
of the dust in the 29~Jan~2010 Hubble Space Telescope (HST) image and
drawing a polygonal fit to them
(Fig.~\ref{fig:conemodelEnvelope}).\footnote{HST images were
  originally taken by \cite{Jew+10}, and were obtained from the {\sc
    STScI} archive}. This procedure is admittedly imprecise and
somewhat arbitrary, but the main envelope drops down sharply to the
apparent sky level, so there is only modest latitude in drawing the
envelope, neglecting the diffuse features (Fig.~\ref{fig:schematic})
in the first part of the analysis.  Furthermore, our purpose is to
perform a simplified fit to show the plausibility of a single simple
burst event creating features similar to those seen in \Atwo, not to
perform the far more difficult task of an exact fit to the outburst.
The main features that we try to explain are the bottom sharp corner
of the dust envelope, $4\farcs4$ south and $3\farcs1$ west of the
nucleus, and a gentler bend $3\farcs6$ to the north and $11\farcs1$ to
the west, with the envelope converging from these bends to a point at
the nucleus and extending in a broad tail to the west.  That is, we
represent the envelope as a five--sided polygon.  Figure
\ref{fig:conemodelEnvelope} shows our best estimate of the envelope
superimposed on the HST image, as well as three other plausible
envelopes to test the sensitivity of our conclusions on the
subjectivity of the envelope.

\subsection {General behavior of the conical burst model}
\label{subsec:generalBehav}

As a first experiment in how debris models can vary in appearance, we
integrate models with fixed $\alpha_c=45^\circ$ and ${{v_{\rm d}}_{\rm
    max}} = 33\,{\rm cm\,s^{-1}}$ in a set of $\phi_{A2}, \theta_{A2}$
directions consistent with the directions of a cube in
$(x_{A2},y_{A2},z_{A2})$ space.  There are $3^3-1=26$ such directions,
made up of all permutations of $(x,y,z)$ with $x,y,z
\in\left\{-1,0,+1\right\}$ excepting $(0,0,0$).  For instance, the
direction $(1,1,1)$ corresponds to $ \phi_{A2}=\theta_{A2}=45^\circ$.

\begin{figure}
\begin{center}
\includegraphics[width=23cm, angle=-90]{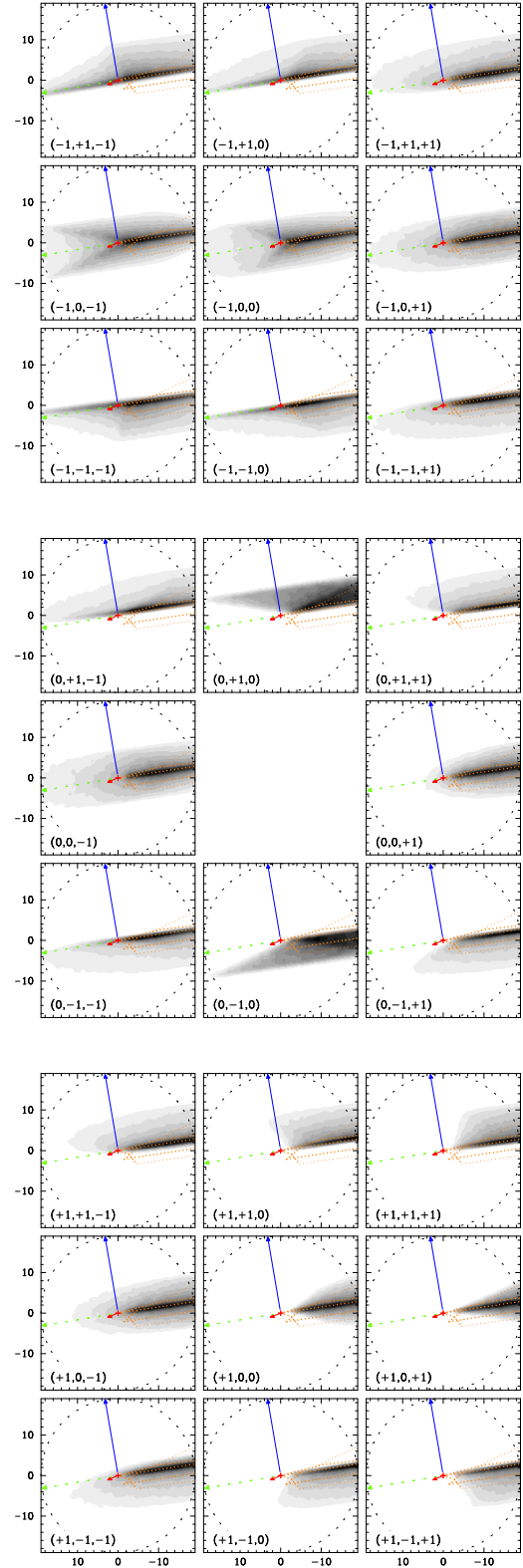}
\caption{Dust simulations of a single conical eruption with \newtext{ fixed
  maximum velocity 
  ${v_{\rm d}}_{\rm max}=33{\rm cm\,s^{-1}}$ and cone angle $\alpha_c=45^\circ$} but with varying emission
direction, as described in Sec.~\ref{subsec:generalBehav}. Units are
arcseconds in RA (abscissa) and Dec (ordinate), in the usual N--up,
E--left orientation.}
\label{fig:multimodel}
\end{center}
\end{figure}

Figure \ref{fig:multimodel} shows this ensemble of models.  In each
sub--plot, the red arrow indicates the projected direction to the
sun, the blue arrow is the vector out of the orbital plane, and the
green arrow is the orbital direction.  Solid (dotted) vectors point
out of (into) the page.  The orange dots correspond to some more
obvious streaks and boundaries of the observed dust distribution;
these are not the dust envelope used in the actual fit in the
subsequent section.

Figure \ref{fig:multimodel} demonstrates that a wide range of
morphologies can be produced by varying only the ejection direction.
The sharp bends that are seen in the actual data are observed in many
of the models, and result from sheet--like or slab--like configuration
of the final dust distribution, with the boundary of the slab
determined by the maximum dust velocity.

\subsection{Fitting procedure and results}

Because small changes in the geometry lead to large changes in the
appearance of the model, and because the penalty function of the fit
is not smoothly varying and may possess multiple minima, we employ a
two--dimensional downhill simplex minimizer described in
\cite{press+92}, running over a grid of the two remaining variables.
Specifically, we freeze $\alpha_c$ and $\phi_{A2}$ over a $10^\circ$
mesh of discrete values, and optimize only in the two dimensions
$\theta_{A2}, {v_{\rm d}}_{\rm max}$.  This particular optimization
scheme is motivated by several considerations: First, ${v_{\rm
    d}}_{\rm max}$ represents a stable and smooth rescaling of
envelope size, and may be safely placed inside the simplex optimizer.
Next, the cone angle $\alpha_c$ is of physical interest, so the method
should produce contours in it.  Finally, the dust envelope varies
unpredictably with the remaining parameters $\phi_{A2}$ and
$\theta_{A2}$, so that not more than one of them should be placed into
the simplex optimizer in order to avoid false local minima.

We find that positive values of $\phi_{A2}$ do not fit the bottom
portion of the envelope at all, successfully filling only the top
portion of the tail, and sometimes explaining the northern bend, as
seen in some panes of Fig.~\ref{fig:multimodel}.
Similarly, negative values of $\phi_{A2}$ with small opening angles
$\alpha_c$ can fit the bottom bend of the envelope, and fill the
bottom part of the tail, but fail to account for the top bend.  Large
($\alpha_c>50^\circ$) opening angles tend to produce east--projecting
spines of dust that are not seen in the actual data.  However, a model
with a cone angle $\alpha_c=40^\circ$ and $\phi_{A2}=-20^\circ$, with
best--fit values ${v_{\rm d}}_{\rm max}=0.2 \,\rm m s^{-1}$ and
$\theta_{A2}=74^\circ$, produces good visual agreement with the dust
envelope, as well as the smallest value of the penalty function.  The
bend in the bright envelope around the ``North diffuse feature'' of
Fig.~\ref{fig:schematic} arises naturally as a consequence of the edge
of the cone, blown backward by radiation pressure.

\begin{figure*}[htbp]
\begin{center}
\includegraphics[scale=0.36]{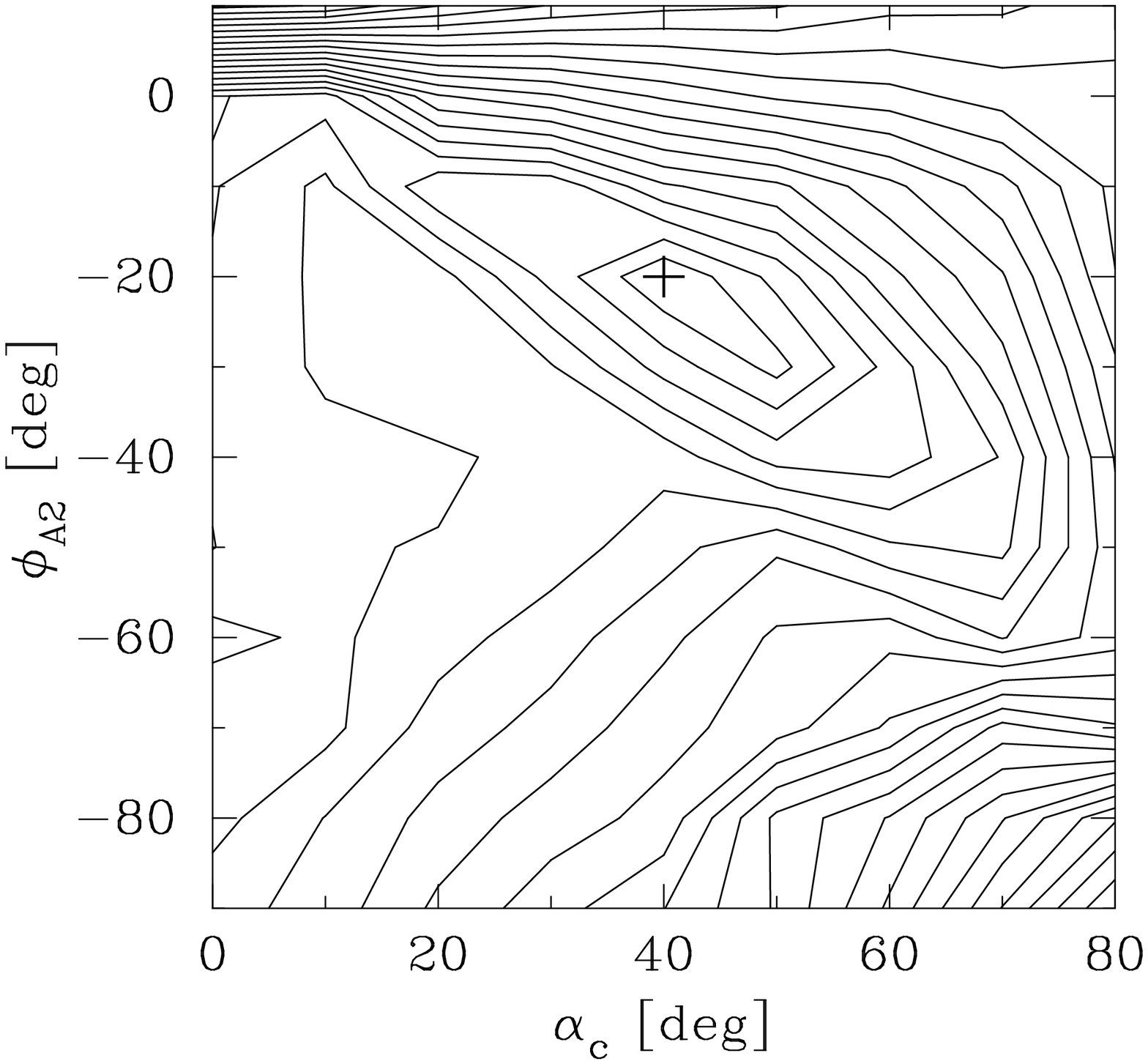} \qquad \quad\includegraphics[scale=0.36]{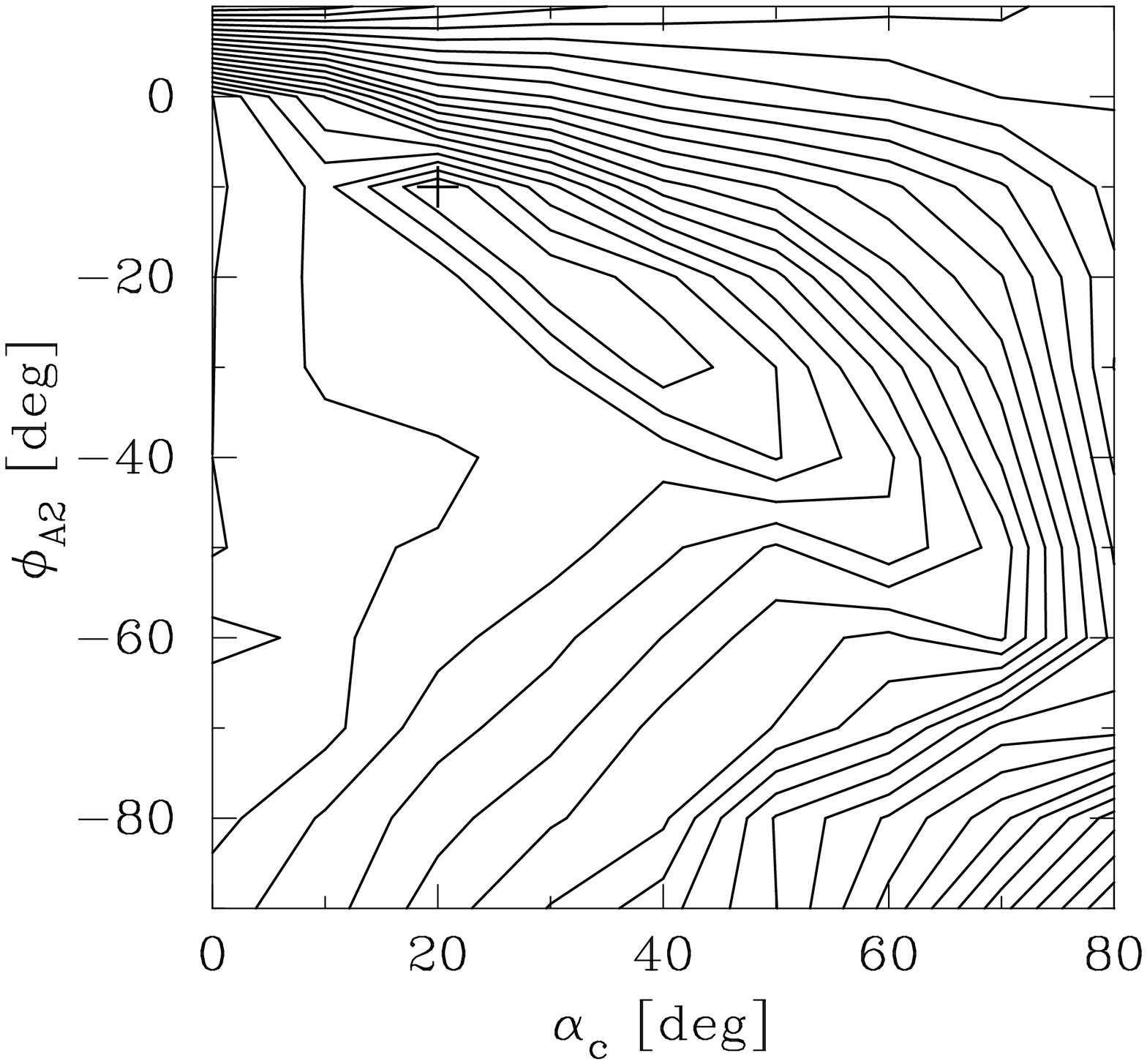} 
\caption{Contours of the binary penalty function in the space of the cone
  opening angle $\alpha_c$ and the latitude direction of ejection
  $\phi_{A2}$, after optimizing over $\theta_{A2}$ and ${v_{\rm
      d}}_{\rm max}$.  The left panel is the fit for the best estimate
  of the envelope, and the right panel is the most discordant (green)
  fit from the envelopes in Figure \ref{fig:conemodelEnvelope}.  Units
  are arbitrary, and there is no statistical interpretation of the
  contour values because they are not $\chi^2$ values. For the best
  envelope, there is a clear optimum at $\phi_{A2}=-20^\circ$ and
  $\alpha_c=40^\circ$, denoted by a plus symbol.  The optimum is
  shifted only slightly for other assumed envelopes. }
\label{fig:conemodelContours}
\end{center}
\end{figure*}

\begin{figure*}[htbp]
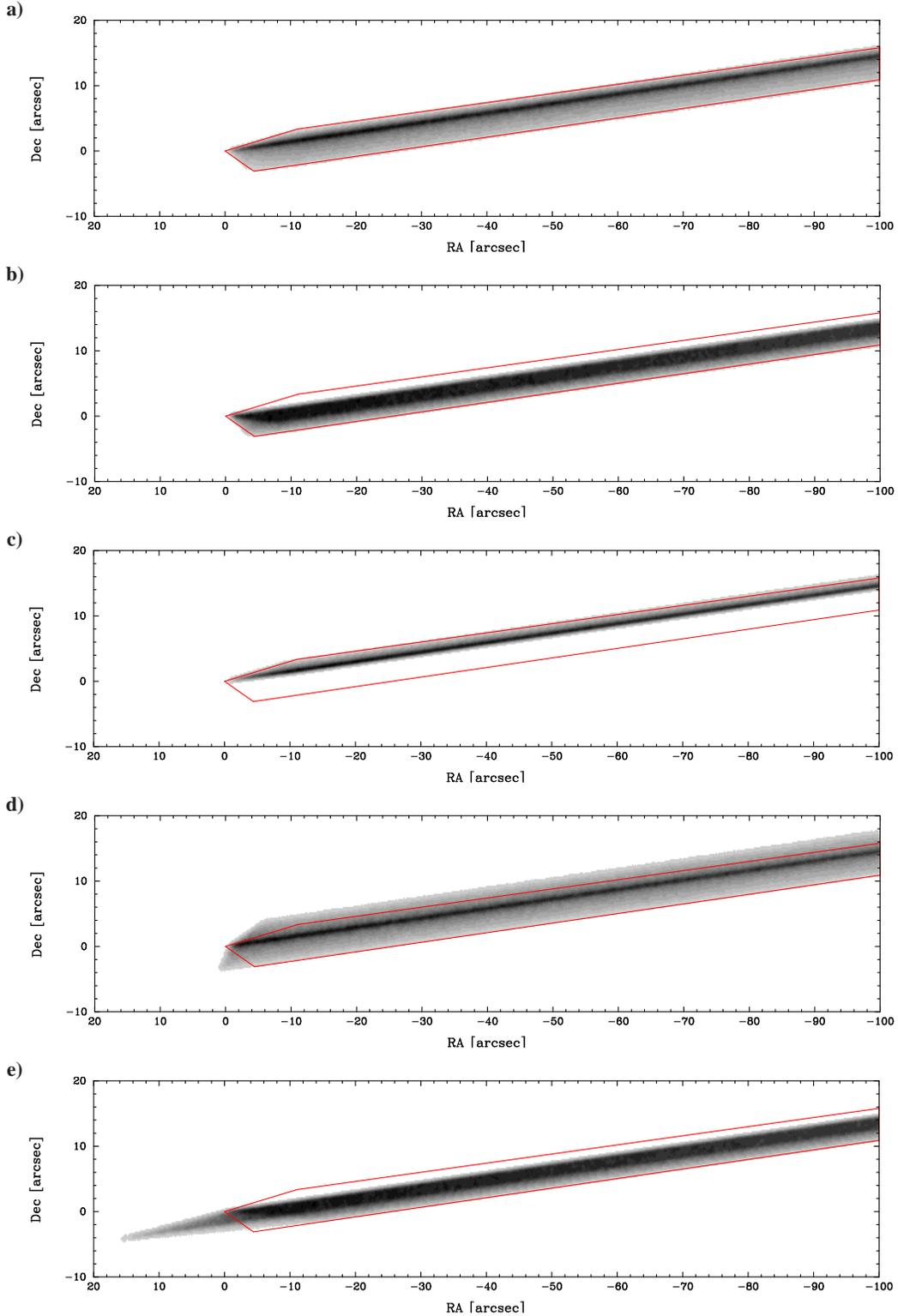

\begin{center}
{\bf a)} \includegraphics[scale=0.55, angle=-90]{burst-fit_hollow-cone_40deg__phi_-20_REVERSED.eps}
\vskip 5pt
{\bf b)} \includegraphics[scale=0.55,  angle=-90]{burst-fit_hollow-cone_10deg__phi_-30_REVERSED.eps}
\vskip 5pt
{\bf c)} \includegraphics[scale=0.55, angle=-90]{burst-fit_hollow-cone_10deg__phi_10_REVERSED.eps}
\vskip 5pt
{\bf d)} \includegraphics[scale=0.55, angle=-90]{burst-fit_hollow-cone_80deg__phi_-20_REVERSED.eps}
\vskip 5pt
{\bf e)} \includegraphics[scale=0.55, angle=-90]{burst-fit_hollow-cone_60deg__phi_-80_REVERSED.eps}
\caption{a) The best hollow cone model, filling the estimated \Atwo\
  dust envelope (red).  This model consisted of a cone with an opening
  angle $\alpha_c=40^\circ$, pointing downward at
  $\phi_{A2}=-20^\circ$ and forward at $\theta_{A2}=74^\circ$, with a
  maximum dust velocity ${v_{\rm d}}_{\rm max}=0.2 \,\rm m
  s^{-1}$. The central streak is the zero--velocity component of the
  dust distribution. b) A hollow cone model with a thin cone
  $\alpha_c=10^\circ$ and $\phi_{A2}=-30^\circ$.  Many thin cone
  models pointing downward successfully fill the bottom part of the
  envelope, but do not account for the top. c) {A hollow cone model
    with a thin cone $\alpha_c=10^\circ$ and $\phi_{A2}=+10^\circ$.
    Models with $\phi_{A2}=>0^\circ$ tend to fill or overfill the top
    part of the envelope, but leave the bottom part empty. 
    \newtext { d) A broad cone with  $\alpha_c=80^\circ$ and
    $\phi_{A2}=-20^\circ$ tends to overfill the envelope, producing a
    spur on the left, and fails to match the top contour of the envelope.}
    e) A broad cone $\alpha_c=60^\circ$ pointing downward at
 $\phi_{A2}=-80^\circ$.  Models with $\phi_{A2}=<-50^\circ$ tend to  produce east projecting dust, and
 very broad cones tend to overfill the envelope at some points in order to fill it at others.}}
\label{fig:conemodels}
\end{center}
\end{figure*}

Figure \ref{fig:conemodelContours} shows penalty function
contours in $\alpha_c$ and $\phi_{A2}$; the contours are very similar
for the various assumed envelopes of Fig.~\ref{fig:conemodelEnvelope}.
Figures \ref{fig:conemodels}a to e
illustrate the best fit model, as well as models that optimize
$\theta_{A2}$ and ${v_{\rm d}}_{\rm max}$, but are away from the
best $\phi_{A2}$ and $\alpha_c$.  Repeating the process using a solid
rather than hollow conical burst gives the same best model, because
the two cone types differ only though the presence of a spherical
instead of planar end--cap.

\begin{figure*}[htbp]
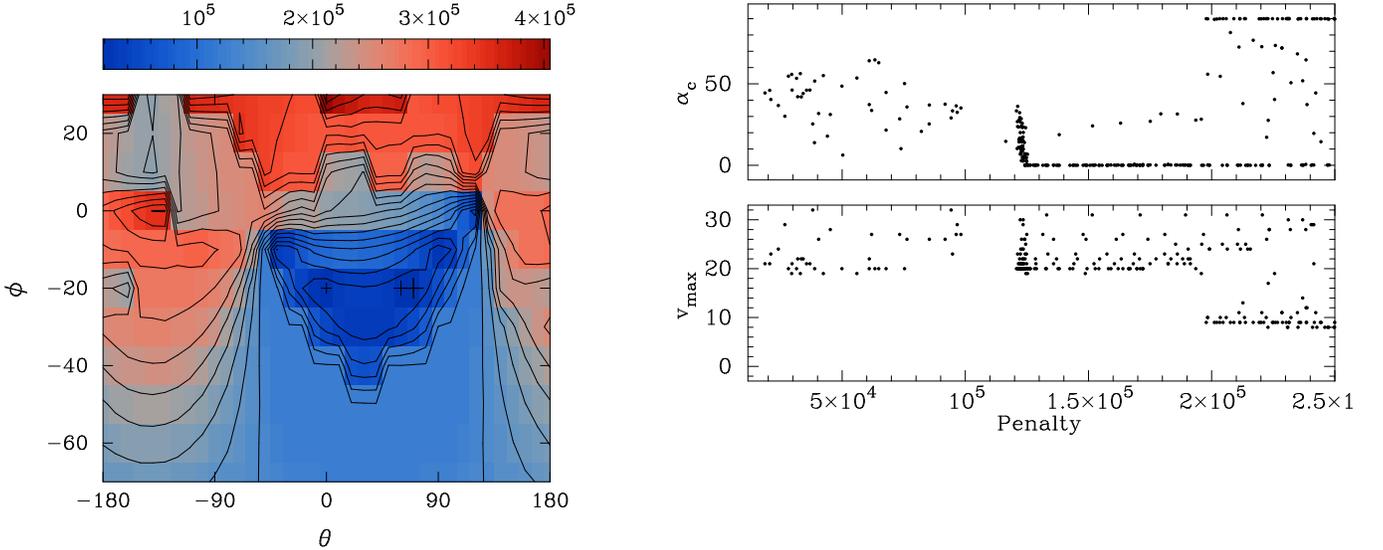

\begin{center}
\includegraphics[scale=0.4,angle=270]{penalty-contours-envelope-method-color.eps} 
\qquad \qquad 
\includegraphics[scale=0.38,angle=270]{fit-vars-vs-penalty-envelope.eps}
\vskip30pt % why is this necessary
\caption{Left: Contours of the binary envelope  penalty function in $\theta, \phi$,
  with the largest plus symbol indicating the minimal value, and the
  other symbols indicating the next two best values.  The best
  solutions are seen to exist in two degenerate islands. Right:  The
   remaining free parameters $\alpha_c$ and $v_{\rm
    max}$ as a function of the penalty.   The two islands of solutions
  give similar values for both parameters, so the choice of solution
  does not affect the physical interpretation of the event, beyond the direction.}
\label{fig:thetaphiminConeModelContours}
\end{center}
\end{figure*}

In addition to performing the minimization by fixing $\alpha_c,
\phi_{A2}$, we also ran an optimization over a grid of fixed
$\theta_{A2},\phi_{A2}$.  Figure
\ref{fig:thetaphiminConeModelContours} shows that there is a second
good fit value at $\theta_{A2}\approx0^\circ,
\phi_{A2}\approx-20^\circ$, with a similar cone opening angle as the
first solution.  Therefore, we must keep in mind that our first
solution is degenerate with a second slightly worse solution.  The two
solutions are related in a physically straightforward manner, evident
upon viewing the model in three dimensions: in one solution, one side
of the cone agrees with the direction of solar pressure, so that
particles in this direction remain on a concentrated stream rather
than being dispersed, and the opposite side of the cone points at the
observer, and is dispersed by sunlight.  In the other solution, the
cone is rotated by nearly $90^\circ$ (two cone half-opening angles),
and debris on the other side of the cone remains coherent, while the
first side points away from the observer is dispersed.  The existence
of two solution differing by roughly a right angle is thus an
independent argument for a $\alpha_c\approx 45^\circ$ cone.  In each
case, the main central streak extending from the nucleus can be
interpreted as the region of the cone over which a spread in
velocities does not result in a transverse (to the stream) spread in
position.  It may be that this alignment helped P/2010 A2 become
visible in the first place, \newtext{ because the brightest features
  of the object appear to coincide with fortuitous line--of--sight
  alignments down the long axis of the dust cone.}

In summary, we found two solutions for the parameters, with
$\theta_{A2}\approx 74^\circ$ or $\sim 0^\circ$, both with a cone half
opening angle of $\alpha_c\approx45^\circ$ and a maximum velocity of $20$ to
$30\,\rm cm\,s^{-1}$.  The two solutions correspond to a rotation of
the cone by its full width.  Because this is not a $\chi^2$ fit,
formal uncertainties cannot be provided. From plots of the
simplex--optimized models on the $\phi_{A2}, \alpha_c$ grid, we
conclude that only models with $30^\circ \lesssim \alpha_c \lesssim
50^\circ$ and $10^\circ \lesssim \phi_{A2} \lesssim 30^\circ$
plausibly fill the dust envelope, providing approximate constraints
for these two parameters.  Models outside the third contour of
Fig.~\ref{fig:conemodelContours} do not appear correct 
\newtext{(e.g.~Fig.~\ref{fig:conemodels}d)}.

% does not work
% \afterpage{\clearpage} %% force outputting of figures so that figs 3 and 4 are not reversed

%%% fixme - should this go later?
% \subsection{Discussion}

% The four dust envelopes considered produced peak dust velocities
% ${v_{\rm d}}_{\rm max}=21 \,\rm cm s^{-1}$ to $35 \,\rm cm
% s^{-1}$.  A spherical body $120\rm m$ in diameter
% \cite{Snodgrass.A2.2010.Nature} with a density of $3\,\rm g \,cm^{-3}$
% is expected to have an escape velocity of $10\,\rm cm\, s^{-1}$,
% nearly an order of magnitude smaller in energy.  Therefore it seems
% impossible for this event to be caused by rotational bursting from the
% YORP effect, because the characteristic velocity of the surface during
% such disruption would be on the order of the escape velocity.  The
% only way to permit a YORP breakup under such circumstances would be to
% assume that the surface is highly cohesive, because all loose dust and
% rubble would already have been hurled free.  

%% the bounding boxes in the following figures were obtained through trial and error

% \begin{figure*}[htbp]
% \begin{center}
% \includegraphics[scale=0.60]{burst-fit_hollow-cone_10deg__phi_-30_REVERSED.eps}
% \caption{A hollow cone model with a thin cone $\alpha_c=10^\circ$ and $\phi_{A2}=-30^\circ$.  Many 
% thin cone models pointing downward successfully fill the bottom part of the envelope, but do not account 
% for the top.}
% \label{fig:conemodelAlpha10PhiMinus30}
% \end{center}
% \end{figure*}

% \begin{figure*}[htbp]
% \begin{center}
% \includegraphics[scale=0.60]{burst-fit_hollow-cone_10deg__phi_10_REVERSED.eps}
% \caption{A hollow cone model with a thin cone $\alpha_c=10^\circ$ and $\phi_{A2}=+10^\circ$.  Models with
% $\phi_{A2}=>0^\circ$ tend to fill or overfill the top part of the envelope, but leave the bottom part empty.}
% \label{fig:conemodeAlpha10PhiPlus10}
% \end{center}
% \end{figure*}

% \begin{figure*}[htbp]
% \begin{center}
% \includegraphics[scale=0.60]{burst-fit_hollow-cone_60deg__phi_-80_REVERSED.eps}
% \caption{A hollow cone model with a broad cone $\alpha_c=60^\circ$ pointing downward at
%  $\phi_{A2}=-80^\circ$.  Models with $\phi_{A2}=<50^\circ$ tend to  produce east projecting dust, and
%  very broad cones tend to overfill the envelope at some points in order to fill it at others.}
% \label{fig:conemodeAlpha60PhiMinus80}
% \end{center}
% \end{figure*}

\section{Modeling the general dust envelope - a multiparametric
  fitting approach}
\label{sec:dustcomplex}

The dust fitting approach of \S\ref{sec:dust} relied on a
hand--drawn trace of the dust envelope, based on the isophotes of the
images.  It possesses the advantages of fitting speed, fit robustness, and
simplicity, but  has the drawback of subjectivity.  To address this
concern, we constructed a more complex model with a non-binary fit,
fitting the observed dust profile, after modelling the emission with a set of
power laws.

Specifically, we assume that the initial differential dust distribution in dust $\beta$
and velocity $v_d$,  before applying an escape
velocity $v_e$, is given by
\begin{equation}
\label{eq:multiparfit}
N(\beta,v_d)\propto 
    \Theta\left(v_0 (\beta / \beta_0)^\delta -v_d \right)
     \times {v_d}^\nu  \beta^\gamma
\end{equation}
In this equation $ \Theta\left( v_0 (\beta / \beta_0)^\delta -
  v_d\right)$ is the Heaviside step function that truncates the
maximum velocity as a powerlaw $(\beta/\beta_0^\delta)$, where
$\beta_0=10^{-5}$ is fixed.  Within this truncation, the dust velocity
is a powerlaw in velocity $N\propto {v_d}^\nu$, and a powerlaw in dust
size $N\propto \beta^\delta$.

It is useful to examine the observational implication of each
parameter: $\gamma$, the dust exponent, determines the fading of the
trail along its long axis.  $\nu$, the velocity exponent, determines
the profile of the trail in the direction of ejection, and its effect is
co--mingled with the effects of geometry and light pressure. $\delta$,
by truncating the velocity in a manner that depends on $\beta$,
controls the broadening of the trail as it extends from the nucleus,
because smaller and more distant dust particles are ejected faster.
$v_0$ sets the limiting envelope of the dust trail.  Finally, the
escape velocity $v_e$ depopulates regions close to the nucleus (and by
extension the center of the trail) by remapping the low velocity dust
distribution according to $v_{\rm final}=\max\left( 0, {v_d}^2 -
  {v_e}^2 \right)^{1/2}$.  Thus this parameterization, even if it does
correspond perfectly to underlying physics, captures many of the
observational variables of the system.

The truncation in velocity is similar to \cite{Jorda2007Icarus}, which
is based on \cite{OKeefeAhrens1985Icarus}, who find that maximum
debris velocity is truncated for a given particle mass as $v_{\rm
  max}\propto m^{-1/2}$ to $m^{-1/6}$ with the best fit being
$m^{-1/3}$.  From the fact that $\beta$ is inversely proportional to
the dust radius, it is true that $\beta\propto m^{-1/3}$, implying
that $\delta\in [0.5, 1.5]$, with $\delta=1$ corresponding to the best
fit of \cite{OKeefeAhrens1985Icarus}.  \footnote{In this paper,
  $\delta$ is the powerlaw index in $\beta$, which differs from the
  nomenclature of \cite{Jorda2007Icarus} and
  \cite{OKeefeAhrens1985Icarus}, where it is the index for particle
  size $a$.}
However, we caution that this form of the truncation is based on
fragmentation assumptions, and it is not obviously applicable to a
loose regolith.  In fact, the value of $\delta$ inferred from the
literature appears inapplicable to this system, because $\beta$ spans
values of $10^{-6}$ to $10^{-4}$ between 0\arcsec and 150\arcsec from
the nucleus, implying that the envelope of the tail must expand by a
factor of at least 10 (for $\delta=0.5$) from the nucleus to the
boundary of our images.  From inspection of the images, the trail
does not appear to broaden by this much, although the broadening
may occur in a faint and invisible halo outside the visible envelope.

For the differential size $a$ distribution of particles, Paper I found
$dN/da\propto a^{-3.5}$, similar to the analytical result for a
relaxed population \citep{Doh69}.  This corresponds to a distribution 
$dN/d\beta \propto \beta^\gamma$ for $\gamma=1.5$.

\cite{OKeefeAhrens1985Icarus} find that the fraction of mass ejected
below a particular velocity $v$ is given approximately by $M/M_{\rm
  Total} \approx 1- v^{-1.3}$, so that the differential exponent
$\nu\approx -2.3$ as long as the particle velocity is independent of
particle size.

 % $\propto \left[1+(R/7\arcmin)^2\right]^{-1}$  

The penalty function of the multi--parameter fit, instead than being
binary as in \S\ref{sec:dust}, uses the absolute difference of the
predicted image with the observed image as the fit metric: ${\rm
  Penalty}=\left| {\rm Image} - {\rm Model} \right|\times
\left(1+(R/R_0)^2\right)^{-\eta}$, where $R$ is the distance from the
nucleus in arcseconds and $R_0$ is a scale parameter.  We use a robust
absoute deviation penalty rather than the more common quadratic one,
because a quadratic fit is strictly correct only for known Gaussian
uncertainties, whereas our images are dominated by unknown
systematics.  Additionally, the penalty is scaled by a
$\left(1+(R/R_0)^2\right)^{-\eta}$ weighting function centered on the
nucleus to suppress the fit at large radii and prevent noise features
in the image from dominating, because most of the image area is
background, far from the nucleus and away from the trail.  The final
fit values are insensitive to $R_0$ between $7\arcsec$ to $30\arcsec$,
and $\eta\in[0.5,1]$, though smaller values of $R_0$ experience more
reliable fit convergence.

To perform the fit, we optimized over a grid of fixed
$\theta_{A2},\phi_{A2}$, and used Powell's method \citep{press+92}
because it is better suited for the smooth powerlaw parameters than
the previous downhill simplex.

\begin{figure*}[htbp]
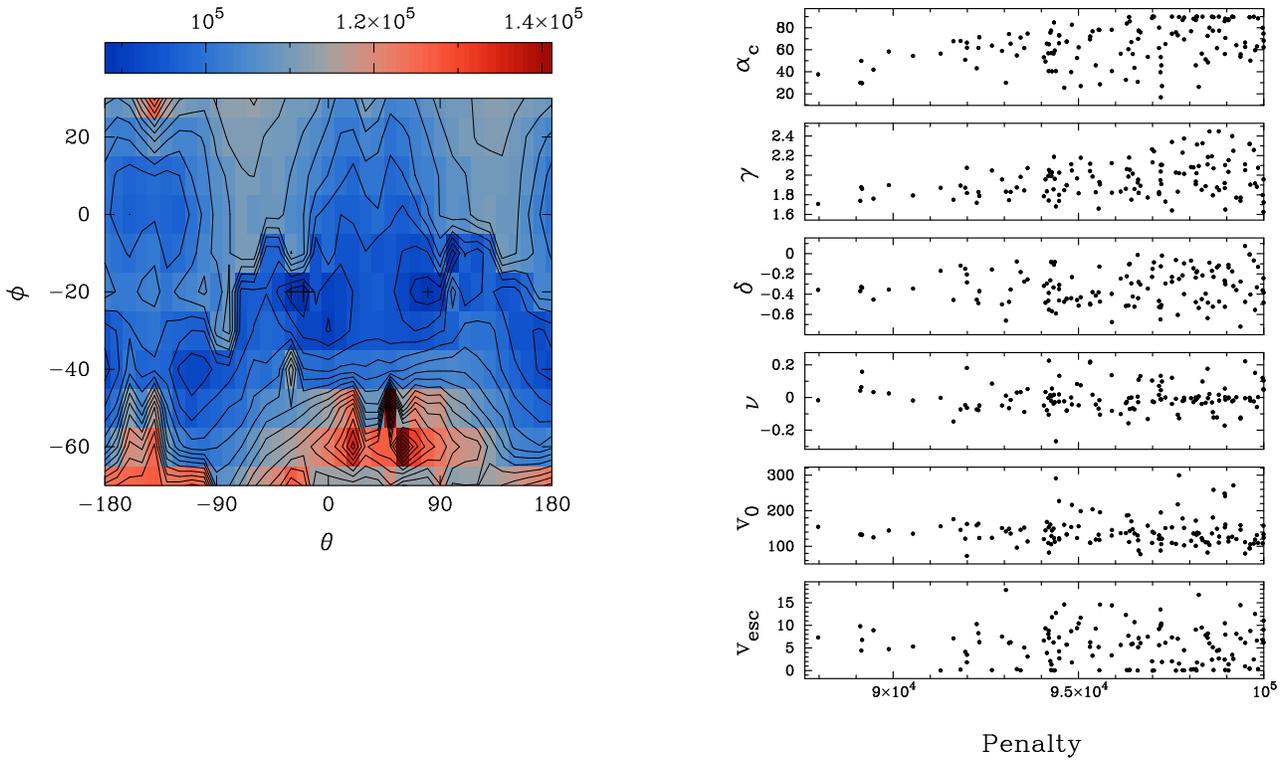

\begin{center}
\includegraphics[scale=0.4,angle=270]{penalty_contours_full_method-color.eps}
\qquad \qquad \qquad
\includegraphics[scale=0.6,angle=270]{fit-vars-vs-penalty-full-fit.eps}
\vskip30pt % why is this necessary
\caption{Left: Contours of the general multiparametric penalty
  function in $\theta, \phi$, with the largest plus symbol indicating
  the minimal value, and the other symbols indicating the next two
  best values.  As for the binary envelope fit, the best fit cone
  solutions are seen to exist in two degenerate islands.  Right: the
  various fit parameters, defined in  \S\ref{sec:dustcomplex}, of
  the multiparametric model as a function of the penalty.  The
  exponents $\gamma, \delta$ and $\nu$ are dimensionless, $\alpha_c$ is in
  degrees, and velocity scale $v_0$ and escape velocity $v_e$
  are in $\rm cm\,s^{-1}$. All of the parameters tend to converge at the
  fit optimum, with the two islands of solutions giving similar
  values. }
\label{fig:thetaphiminConeModelContoursGeneral}
\end{center}
\end{figure*}

Figure \ref{fig:thetaphiminConeModelContoursGeneral} (left) shows the
resulting penalty contours in $\theta_{A2},\phi_{A2}$.  Although not
as clean as in the binary fitting case, the two islands of optimal
solutions are at the same location as before, with the optimal island
now at $\theta_{A2}\approx-10^\circ$ rather than than
$\theta_{A2}\approx80^\circ$.  This agreement suggests that the principal
finding of the ejection direction and opening angle is robust with
respect to the choice of fitting method.

By plotting the other parameters of multi-parametric model against the
fit value, we ascertain whether they converge to consistent values
(Figure \ref{fig:thetaphiminConeModelContoursGeneral}, right).  As in
the case of the binary fit, the opening angle of the cone converges to
$\alpha_c \approx 40^\circ$.  The dust distribution is $dN/d\beta \propto
\beta^{1.7}$, close to the $\beta^{1.5}$ from collision theory
and the fit of Paper I.  Finally, the escape velocity converges to
$v_e \approx 6\, \rm cm\, s^{-1}$, close to the value of $\sim
10 \rm cm\, s^{-1}$ estimated in Paper I on other grounds.  

\begin{figure*}[htbp]
\begin{center}
\includegraphics[scale=0.40,angle=0]{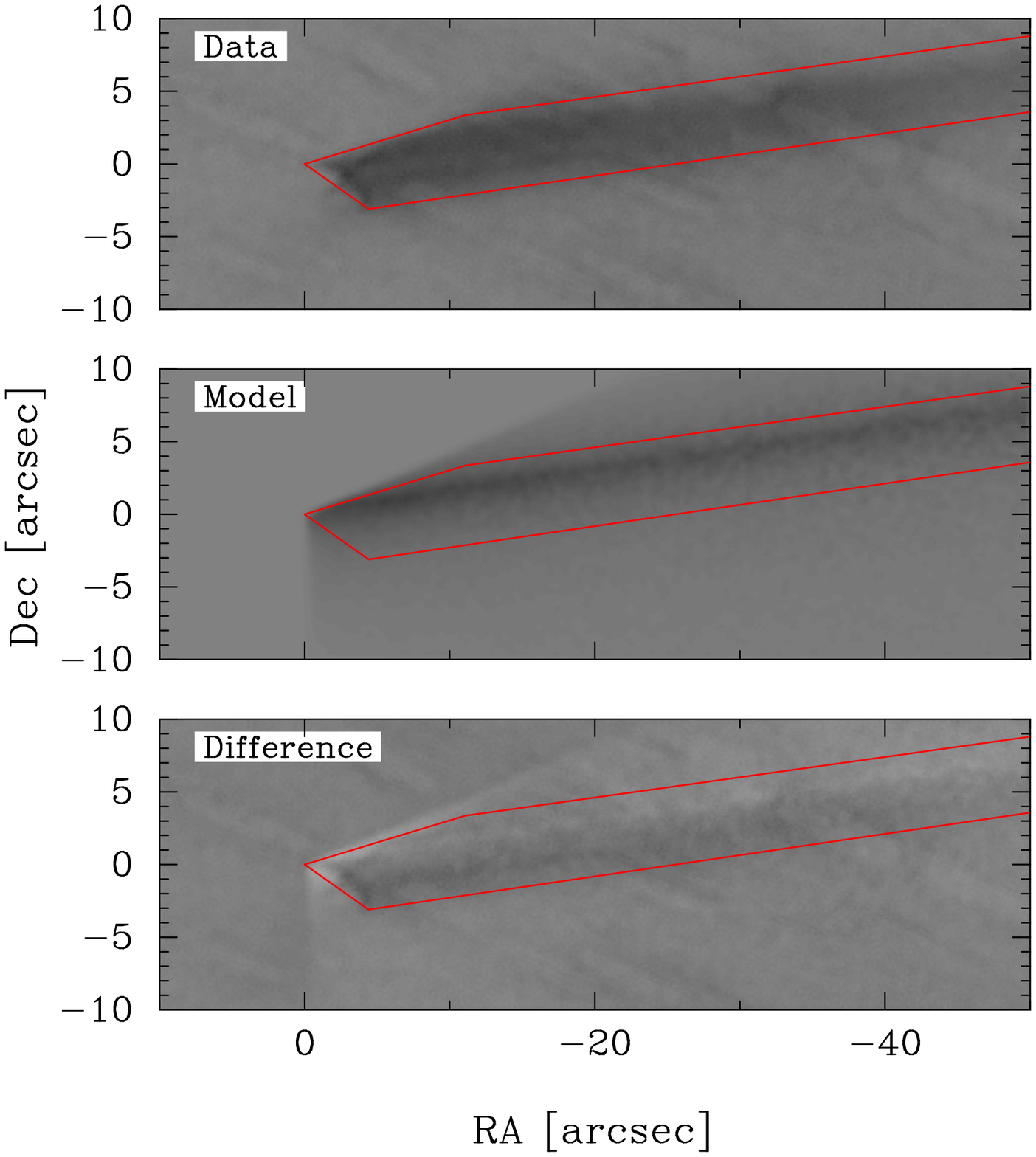}
%\qquad \qquad 
\includegraphics[scale=0.40,angle=0]{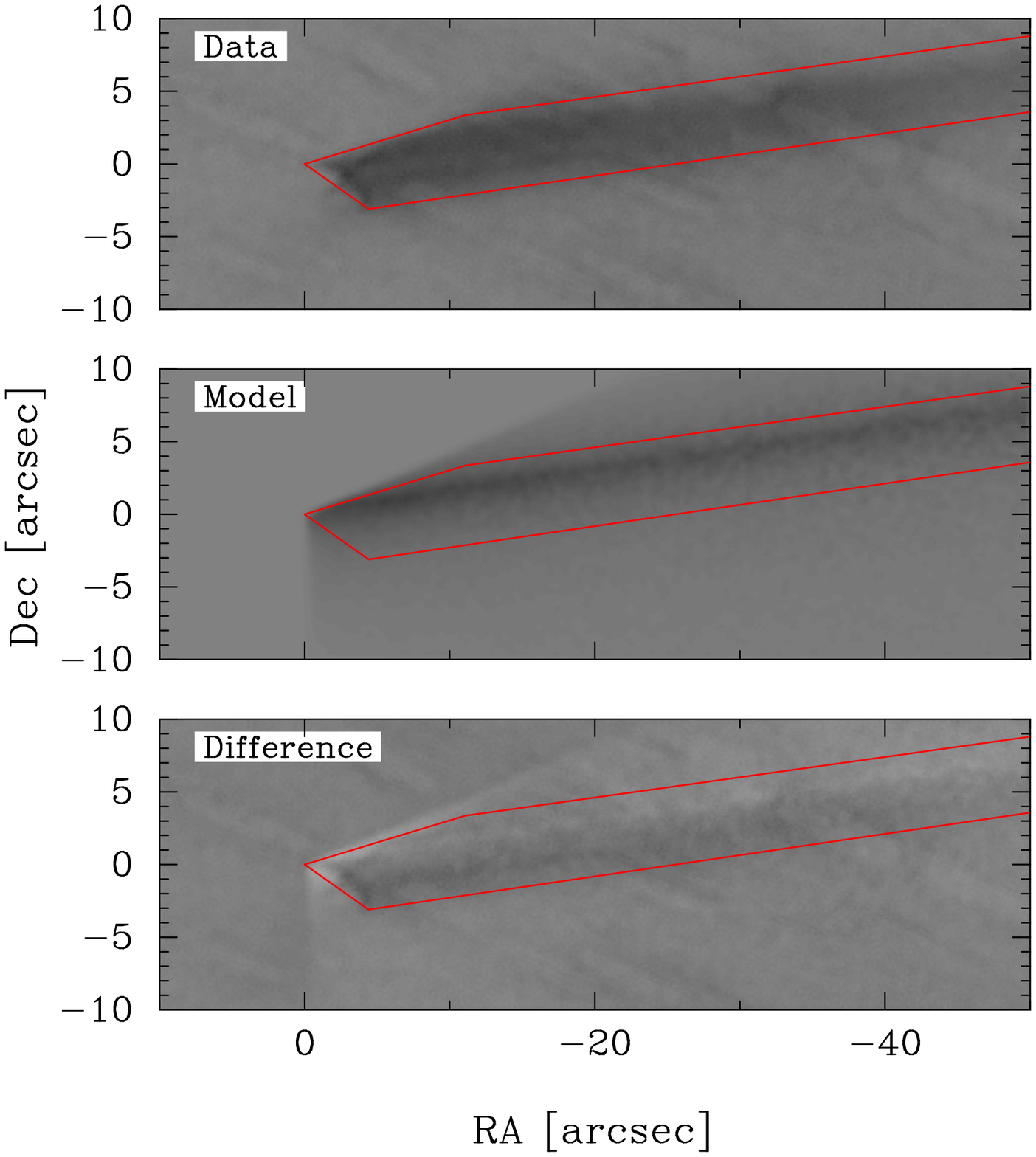}
\caption{Left: Images of a fine-tuned solution at the left solution
  island of Fig.~\ref{fig:thetaphiminConeModelContoursGeneral},
  converging to $\theta_{A2}=-11^\circ, \phi_{A2}=-22^\circ ,
  \alpha_{A2}=45^\circ$.  Right: Images of the the solution with the
  solver started at the right island, converging to
  $\theta_{A2}=79^\circ, \phi_{A2}=-17^\circ, \alpha_{A2}=45^\circ$.
  In each case, the original HST image is in the top panel, the model
  is in the middle, and the image minus the model is in the bottom.
  The envelopes from the previous section are over-plotted for
  reference only.  Unlike the binary fit, this fit incorporates the
  faint diffuse features (Fig.~\ref{fig:schematic}), accounting for the
  higher maximum velocity of $v_0\sim 100\,\rm cm\,s^{-1}$.}
\label{fig:FineTuneMultiparImages}
\end{center}
\end{figure*}

The value $v_0$, representing the cutoff velocity for a
$\beta=\beta_0=10^{-5}$ dust grain, is about $100\,\rm cm\,s^{-1}$,
much higher than the $25 \,\rm cm\,s^{-1}$ binary fit value.  As will
be seen in Fig.~\ref{fig:FineTuneMultiparImages}, this higher velocity limit
results from the fact that the model now fits the ``North diffuse
feature'' and ``SE diffuse Feature'' of Fig.~\ref{fig:schematic}, rather
than the brightest part of the envelope as in the binary fit.

The exponent $\delta$ controlling the cutoff of velocity in
particle size (Equation \ref{eq:multiparfit}) is significantly
different from what is predicted: we obtain $\delta\approx -0.4$ when
previous semi-empirical studies \citep{OKeefeAhrens1985Icarus} suggest
$\delta\in [0.5, 1.5]$.  Recovering $\delta$ depends on measuring the
broadening of the tail far from the nucleus (where $\beta$ is larger)
relative to the width near the nucleus.  $\delta$ is constrained by
the maximum extent of the very faint of the N and SE diffuse
features, which fade rapidly into the background.

Finally, the velocity exponent in $v^\nu$ converges to $\nu\approx 0$,
flatter than the $\nu\approx -2.3$ value predicted by experiment and
simple cratering theory, so that our fit contains more fast--moving
dust than expected from a naive powerlaw model.

Fig.~\ref{fig:FineTuneMultiparImages} shows the appearance of fine-tuned
solutions, starting the optimizer in the two solution islands.  The two solutions at
$\theta_{A2}=-11^\circ$ and $79^\circ$ appear similar.  Both attempt
to account for the N and SE diffuse features, placing some dust in
an approximately correct location, though with large residuals.  The
models underestimate the dust at the bottom of the envelope, in
particular the lower ``striae'' of Fig.~\ref{fig:schematic}.  Such
features are almost certainly not in agreement with the simple
powerlaw assumptions.   Though these models account for
the overall geometry, complex substructures are beyond their scope.

In summary, our second fitting approach considered a model with
powerlaw dependancies on dust size and velocity.  We recovered
approximately the same two islands of $\alpha_c=45^\circ$ solutions as
the simpler binary fit method, this time at $\theta_{A2}=-11^\circ,
\phi_{A2}=-22$ and at $\theta_{A2}=79^\circ, \phi_{A2}=-17^\circ$.
These are again separated by about $90^\circ$ on the sky and differ by
a rotation of one full cone width.  We recover the
previously obtained $\sim a^{-3.5}$ dust distribution, but the
distribution in velocity is flatter than expected from cratering
theory, with more dust at high velocities.  The higher limiting debris
velocity of the powerlaw fit arises from a filling of the N
and SE diffuse features in Fig.~\ref{fig:schematic}.

%\clearpage

\section{Interpretation of the arc features}
\label{sec:arcs}

\subsection{Description of features}

\begin{figure*}[htbp]
\begin{center}
\includegraphics[width=18cm]{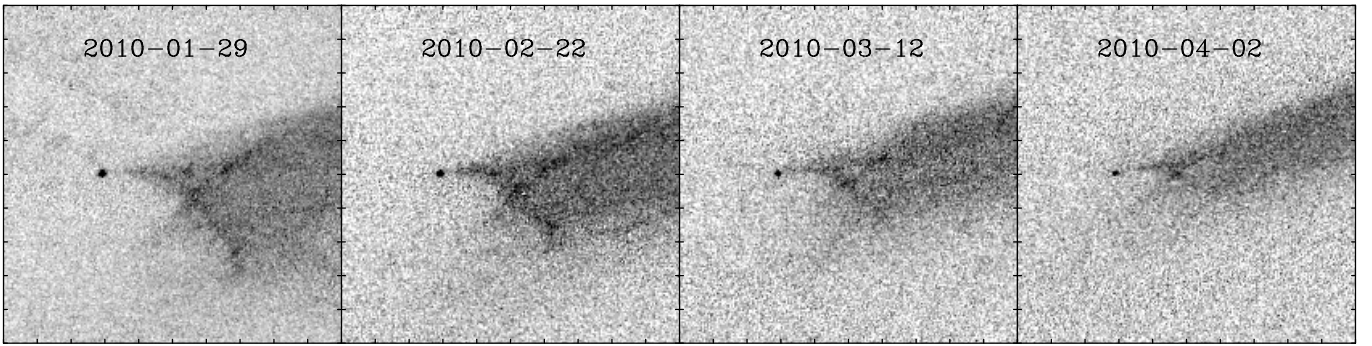} 
\caption{\label{fig:hst4} HST images of \Atwo, north up and east
  left, showing the evolution of the central debris feature with
  time. Each image is $10\arcsec\times10\arcsec$, and
  each tick mark is $1\arcsec$.  Most notable is the cross
  composed of a streak extending southwest from the nucleus,
  intersected by another streak extending to the northeast, denoted as
  Arcs 1 and 2 in Fig.~\ref{fig:schematic}.}
\end{center}
\end{figure*}

To the west of the nucleus, there is a distinctive feature consisting
of what appears to be two crossed trails or arcs
(Fig.~\ref{fig:schematic}), about $10\arcsec$ in size; it is
most visible in the early HST images of Jan and Feb 2010, and traces
of it remain in later months (Fig.~\ref{fig:hst4}).  Unless noted, we
will focus on the feature as it appears on 29 Jan, when it was clearly
traceable.  Arc 2 points at the nucleus to its NE, but
Arc 1 does not contact the nucleus at all, even if extended.
Both features are within the dust envelope employed in the dust
models, although Arc 2, the feature that points downward, is close to the left
boundary of the envelope.

\newtext{Additionally, there are other features like the striae along
  the tail direction, as noted in Fig.~\ref{fig:schematic}; these
  probably arise from secondary dust emission by debris.  These
  features are beyond the scope of our models because there is
  no simple way to parameterize them.  Indeed, these features involve
time dependent processes, whereas our models all assume instantaneous 
ejection. }

\subsection{Possible distributions of debris emission}
\label{subsec:arcInference}

To interpret these arc--shaped, features, we will attempt to use
general arguments of their dimensionality in physical space to
constrain the particle size and the initial directional and velocity
distribution of the debris. 

Assuming a single time of outburst, any distribution of particles in
space arises from a mapping of emission direction and velocity to
three and two dimensional distributions $F_{3D}$ and $F_{2D}$ of the
form
\newtext{ (with the arrow ``$\longrightarrow$'' indicating a mapping from one space to another):}
\begin{equation}
  f(\theta_{A2},\phi_{A2}, v_{\rm d}, \beta) \longrightarrow F_{3D}(x,y,z) 
  \longrightarrow F_{2D}(X,Y) 
\end{equation}
\noindent where $x,y,z$ are three dimensional spatial coordinates, and
$X,Y$ are projected sky coordinates.  In general, an $n$--dimensional
manifold in $\theta_{A2},\phi_{A2}, v_{\rm d}, \beta$
emission-space will map to an $n-$dimensional manifold in $x,y,z$ or
projected $X,Y$ physical space. For example, a two--dimensional
surface in emission--space is described by two internal parameters,
and it will continue to be described by these two parameters when
emission--space is mapped to $x,y,z$ physical space.  In order for
$F_{2D}(X,Y)$ to be a one--dimensional curve, $F_{3D}(x,y,z)$ must be
a one--dimensional curve or a two--dimensional sheet, and the same
must be true of $ f(\theta_{A2},\phi_{A2}, v_{\rm d}, \beta)$.

We first consider the possibility of a range of dust sizes by a simple
numerical experiment: we launch particles at a fine grid in
$\theta_{A2},\phi_{A2}, v_{\rm d}$ in pairs of $\beta=0$ and
$\beta=10^{-5}$. We trace the angular separation of the two valus of
$\beta$ in each pair to determine the dust trail direction for that
launch vector.  We find that the position angle on the sky is always
in a $3^\circ$ range around $278^\circ$, like the main trail.  Hence
there is no configuration of particles that could explain the
cross-feature as a consequence of radiation--pressure driven trailing,
because the position angle of both features is far from $278^\circ$.
{\it Hence the features must consist of a single particle size, and
  the only natural size is very large particles with $\beta=0$.}

Noting that the arcs appear to be one--dimensional curves, there are
two possibilities for the initial emission. First, the emission may be a
one--dimensional curve in the space of $\theta_{A2},\phi_{A2}, v_{\rm
  d}$.  Examples might be a burst of debris in one direction
$\theta_{A2},\phi_{A2}$ over a range of velocities, or emission over a
curve in $\theta_{A2},\phi_{A2}$ at a single velocity. Second, the
emission could be a two--dimensional surface in
$\theta_{A2},\phi_{A2}, v_{\rm d}$, mapping onto a surface in $x,y,z$
that is projected into a thin curve in $X,Y$.  Examples of this second
case might be a fan--like eruption, or one region of a hollow conical
eruption.

\subsection{Orbit models of the features}

\begin{figure}
\begin{center}
\vskip -1cm
\includegraphics[width=8cm, angle=0]{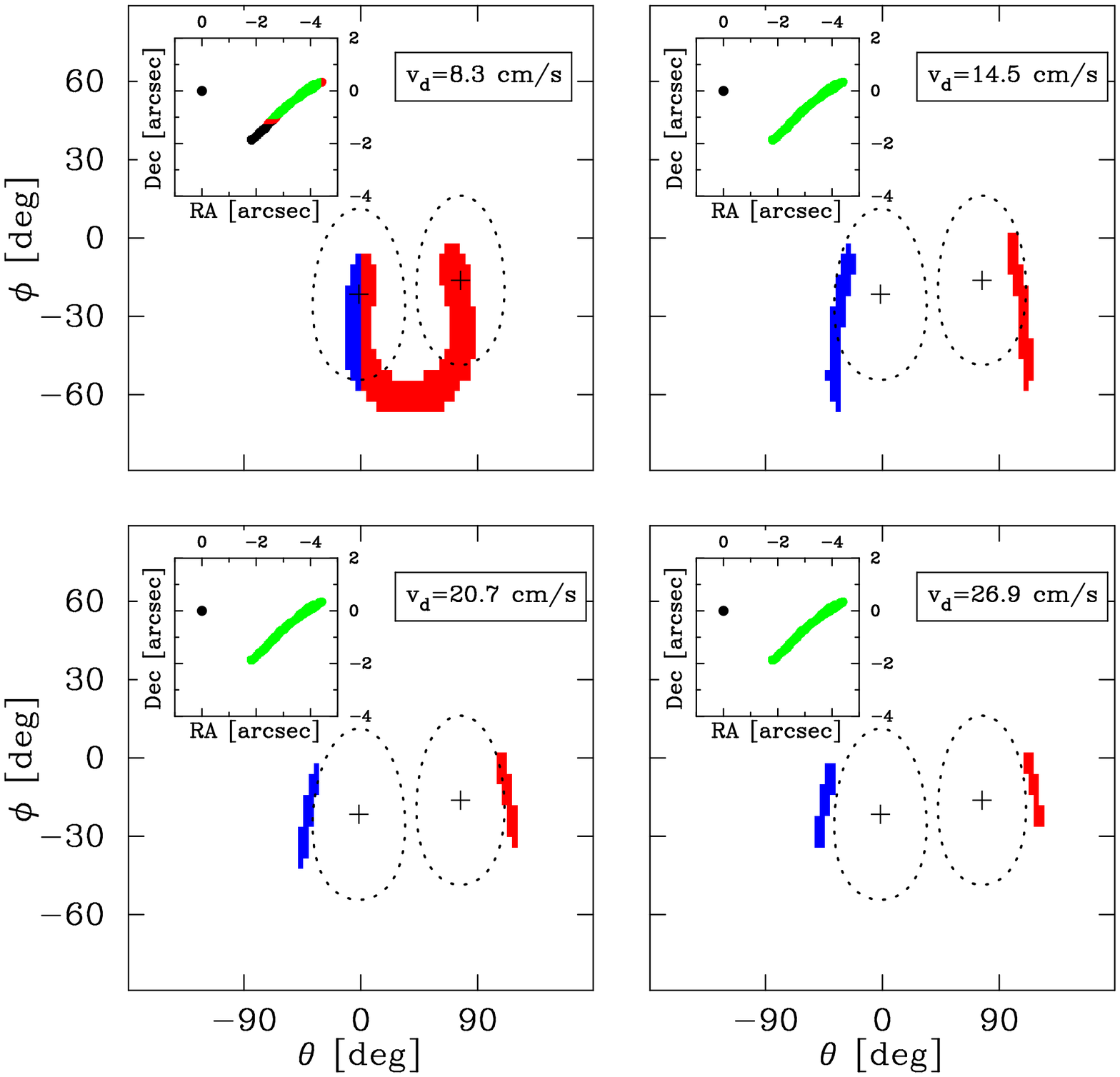} \\
\vskip -2cm
\includegraphics[width=8cm, angle=0]{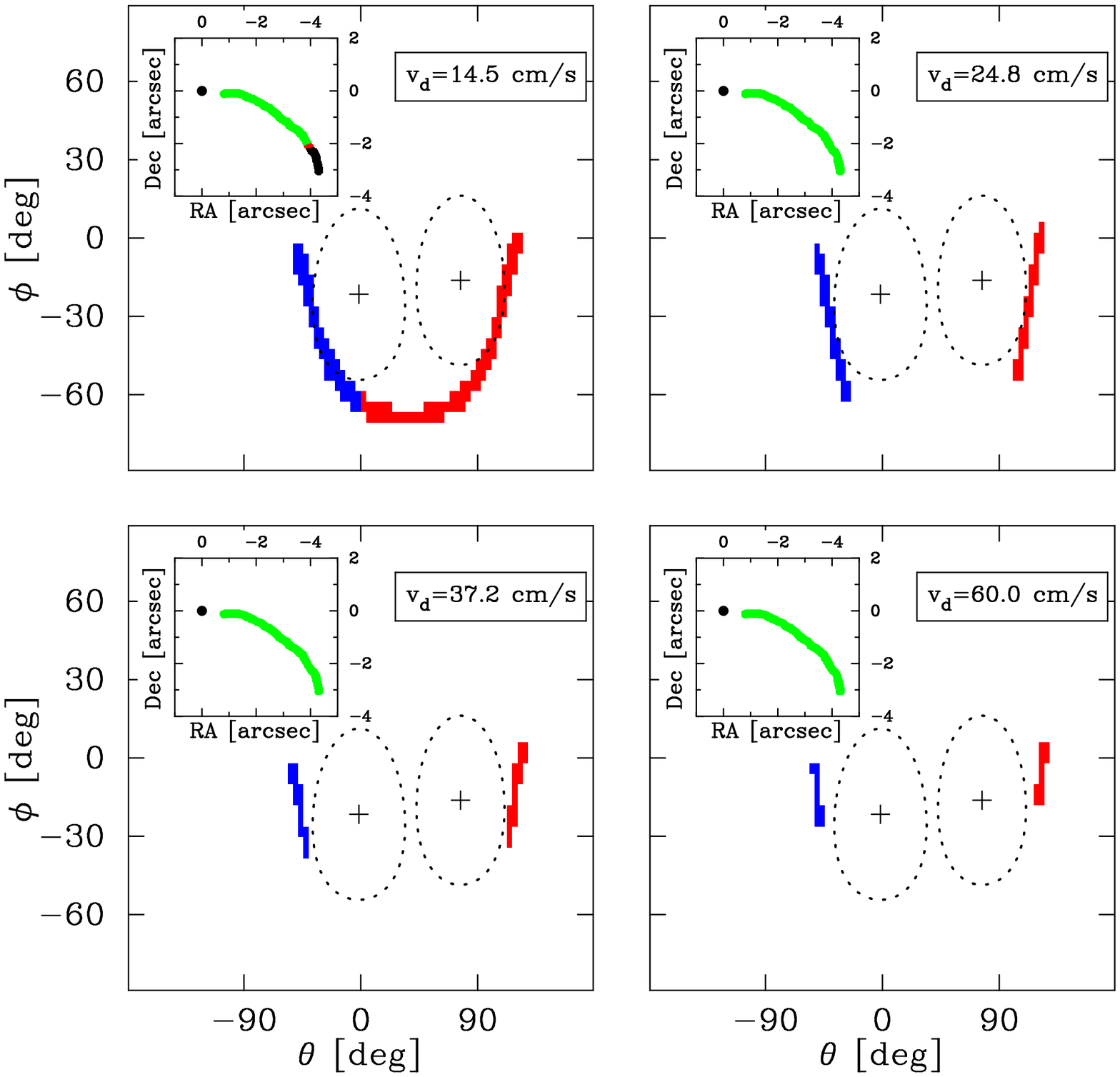}
\caption{\small \label{fig:crossOrbits} A depiction of which large
  particle ($\beta=0$) initial debris orbits emitted on 2009~Feb~10
  and ending on 2010~Jan~29 could have accounted for the Arc 1 (top
  panel) and Arc 2 (bottom panel) features (see
  Fig.~\ref{fig:schematic}). The separate four--panel plots show the
  two components of the cross individually, depicted on the RA, Dec
  plane within the inset.  Within each plot, each of the four panels
  represents one initial particle velocity, and the axes
  $\phi_{A2},\theta_{A2}$ are the emission direction of the debris.
  Colored regions in the $\phi_{A2},\theta_{A2}$ plane are debris
  orbits that fall onto the observed feature; they are coded blue and
  red to denote the two separate islands of solutions on the
  $\phi_{A2},\theta_{A2}$ plane.  In the RA, Dec inset showing the
  feature on the sky, regions of the feature are colored red and blue
  if they are explained by the corresponding orbits in the larger
  $\phi_{A2},\theta_{A2}$ plot, green if both the red and green
  solutions can explain them, and black if no particle orbits can
  explain this part of the feature.  The $+$ symbols and the dotted
  curves around it are the two solutions best fit dust cones from the
  first part of the modeling
  (Fig.~\ref{fig:thetaphiminConeModelContours} and
  \ref{fig:thetaphiminConeModelContoursGeneral} ).  As discussed in
  the text, there is considerably overlap between the edges of the
  best fit dust cone and the discrete features in the insets.  There
  is also a correspondence of the features with the regions of maximum
  foreshortening shown in Figure \ref{fig:orbitProjection}.}
\end{center}
\end{figure}

Having argued that the arcs consist of very large particles with
$\beta=0$, and that the emission is either a one--dimensional curve or
a two--dimensional surface in the emission space
$\theta_{A2},\phi_{A2}, v_{\rm d}$, we next attempt to determine which
directions of emission could account for the feature.  We integrate a
library of particle orbits starting on 10~Feb~2009 and ending on
29~Jan~2010, each orbit defined by its launch direction
$\phi_{A2},\theta_{A2}$ and its velocity $v_d$.  In Figure
\ref{fig:crossOrbits} we show which of these orbits could account for
the two arc components.  Specifically, at a particular $v_d$, we color
those regions of emission $\theta_{A2},\phi_{A2}$ that could populate
each feature (shown in $X={\rm RA}, Y={\rm Dec}$ sky space as an
inset). By stacking these $\theta_{A2},\phi_{A2}$ slices in $v_d$, one
can imagine a colored surface defined on a cube of $\theta_{A2},\phi_{A2},
v_{\rm d}$. Emission must occur on a curve or sub--surface on this
surface in order to fall onto the features.

Figure \ref{fig:crossOrbits} immediately shows that the two
sets of degenerate orbital features (blue and red colored islands)
correspond in some way to the two degenerate dust cones that we
derived previously, using either the binary or multiparametric fitting
method (Fig.~\ref{fig:thetaphiminConeModelContours} and
\ref{fig:thetaphiminConeModelContoursGeneral}).  Apparently, the
viewing geometry is unable to distinguish between two equally good
emission directions, but these two directions do not differ in their
physical implications.

It is evident that there exists a debris velocity $v_d$ below which
the features cannot be explained because the dust cannot reach the arc
position.  This is seen in those velocity slices (panes) of Fig.~\ref{fig:crossOrbits} where the $x,y$
inset picture of the feature is \newtext{ left black rather than red, blue,
  or green, meaning no orbits at that velocity fall on the feature.
  For Arc 1 in
the top four--pane panel of Fig.~\ref{fig:crossOrbits}, the $8.3\, \rm
cm\,s^{-1}$ slice is ruled out, and for Arc 2 in the bottom panel,  the  $14.5\, \rm
cm\,s^{-1}$ slice is excluded. } After this minimum velocity is surpassed, the feature
can be explained only with a rather complicated curve in the
$\phi_{A2},\theta_{A2}$ plane.  However, at larger velocities, the
colored orbits split into two islands on the $\phi_{A2},\theta_{A2}$
plane, each of which suffices alone to explain the feature, again
reflecting the existence of two solutions. 

From the first part of Figure \ref{fig:crossOrbits}, Arc 1 can be
explained by orbits that lie on the edge of the two degenerate
dust cone solutions (dotted curves) that were fit previously.  The
velocities that work are 20 to 30 $\rm cm\,s^{-1}$, consistent with
the dust cone fit.  Because only orbits in a particular velocity range
lie directly on the cones, it appears likely that the feature is
localized in velocity if indeed it is part of the main cone.
Encouragingly, {\it this feature was not a part of the dust envelope
  used in the dust fit} - hence the coincidence of the orbits with the
outline of the dust cone is completely independent of the dust trail
fit.

\begin{figure}
\begin{center}
\vskip -1cm
\includegraphics[width=8cm, angle=0]{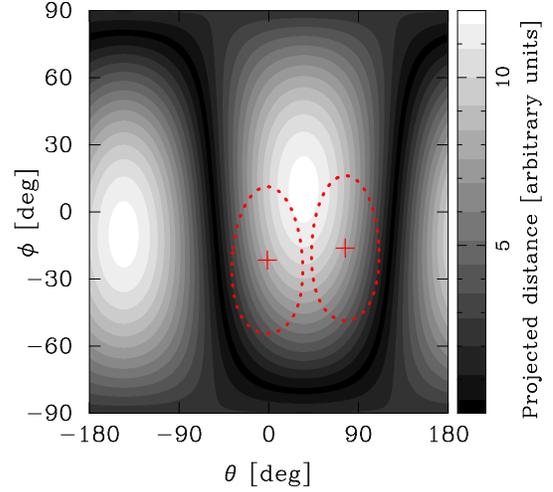} 
\caption{\label{fig:orbitProjection} A depiction of the foreshortening
  of debris orbits emitted on 2009~Feb~10 and ending at 2010~Jan~29,
  with the emission occurring in the direction $\phi_{A2},\theta_{A2}$
  and a velocity $v_d=31 \, \rm cm\,s^{-1}$.  Dark regions indicate
  directions in which particles emitted at slightly different
  velocities remain close on the sky, and light regions are where such
  particles appear more separated.  Strings of debris emitted in
  directions where the figure is dark will be less dispersed and more
  visible.  The red $+$ symbols and the dotted curve around it are the
  nearly degenerate best fit dust cone solutions from the first part
  of the modeling.  The right side ($\theta_{A2}\approx 110^\circ$) of
  the the right cone or the left side of the left cone should have
  enhanced visibility, and, indeed, these regions correspond to the
  two solutions for Arc 1 and 2 in Fig.~\ref{fig:crossOrbits}.
  In particular, the orbits that could account for Arc 2 exist
  exclusively in the darkest regions of this figure.}
\end{center}
\end{figure}

We then address the question of why this particular set of directions
on the dust cone appear as a discrete feature.  Figure
\ref{fig:orbitProjection} shows which emission directions were subject
to greatest foreshortening (and thus visual enhancement) on
29~Jan~2010.  To make this figure, pairs of particles separated by a
$2\,\rm cm\,s^{-1}$ difference in velocity were launched in each
direction at $v_d=28$ and $30\, \rm cm\,s^{-1}$, and the final
separation on the sky within each pair was plotted as a grayscale
image.  Dark regions correspond to emission directions that are
maximally foreshortened.  It is apparent that the parts of the dust
cone (red dotted line) that correspond to the first cross feature are
among those that are most parallel to the line of sight.  Thus the
picture that suggests itself is that Arc 1 is a line of sight
projection of large particles within a narrow velocity range on the
best fit dust cone.  An important caveat with Figure
\ref{fig:orbitProjection} is that it addresses only projection arising
from a velocity spread.  Other types of projection may also
contribute, like the edge brightening of a hollow cone viewed from the
side.

The orbits explaining Arc 2, in the second panel of Figure
\ref{fig:crossOrbits}, also touch the best--fit dust envelope cone,
like Arc 1.  However, these orbits do not fit the cone as well as
those of Arc 1, though the highlighted arc of orbits does agree strikingly
well with the most foreshortened regions in Figure
\ref{fig:orbitProjection}.  Hence it is likely that Arc 2 represents
extended debris from main cone that also happens to fall into
the directions of maximum line--of--sight projection.

When Fig.~\ref{fig:orbitProjection} is recomputed using the features
as observed on 12 March 2010 and 2 April 2010 (see Fig.~\ref{fig:hst4}),
using the analogous but highly evolved features on these dates, the
figure resembles the original 29~Jan~2010 version, demonstrating that
the orbital interpretation remains consistent with time.   

It is also possible to show that both Arcs 1 and 2 may be fully
recreated using only those orbits lying precisely on either of the
cones, by selecting orbits on particular curves in the space of
velocity and the cone's azimuthal parameter. Given the freedom to
place arbitrary curves of overdensity on the depicted cones, one can
easily recreate the features.  However, different cones work as well, and
such {\it ad hoc} curves have no obvious physical interpretation
beyond arising from inherent asymmetries in the ejection event.

In summary, our procedure of mapping out which large--particle debris
orbits could account for the Arcs 1 and 2 shows that the allowed
initial trajectories constitute a surface in emission space, as
surmised in Sec.~\ref{subsec:arcInference}.  Moreover, the
trajectories largely coincide with the previously determined best--fit
dust cone.  For the case of Arc 1 at least, the permitted orbits are
completely independent of the data that went into the fitting of the
dust cone, so there are two separate pieces of evidence pointing at
conical emission in the directions computed.

\subsection{A simple model of \Atwo}
\label{sec:simplemodel}

\begin{figure*}[htb]
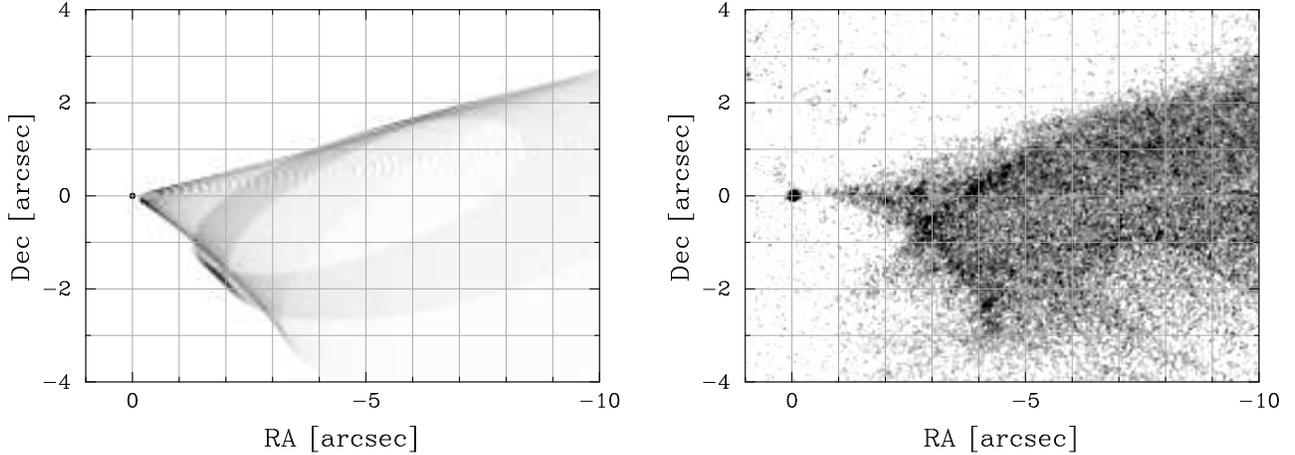

\begin{center}
\includegraphics[width=6cm, angle=-90]{simple-model-final.eps}\qquad
\includegraphics[width=6cm, angle=-90]{tailored-simimg-hst_jan29.eps}
\caption{\label{fig:toyconefeatures}Left: A toy model of $\beta=0$
  debris ejection at $\phi_{A2}=-20^\circ,\theta_{A2}=74^\circ$.  An
  excess of ejecta at a velocity of $14$ to $22\,\rm cm \,s^{-1}$ has been
  used to create the elliptical band inside the cone, resembling one
  of the features of the data.  To illustrate a mechanism by which 
the oberved curvature might be produced, the cone tapers from a width of
  $\alpha_c=40^\circ$ at its outer extreme velocity of $35,\rm cm
  \,s^{-1}$ to a width of $\alpha_c=33^\circ$ at its inner velocity of
  $0\,\rm cm \,s^{-1}$.  .  The boundaries
  of the cone are brightened by edge effects, and correspond very
  roughly to the other main feature of the data.  Right: the
  29~Jan~2010 HST image, for comparison.  The model for the second (degenerate)
  solution at $\theta_{A2}\approx-5^\circ$ appears similar.}
\end{center}
\end{figure*}

Figure \ref{fig:toyconefeatures} illustrates a very simple model based on
some of the inferences we have described.  The principal feature of
this model is a $\beta=0$ dust cone at
$\phi_{A2}=-20^\circ,\theta_{A2}=74^\circ$, with ${v_{\rm d}}_{\rm
  max} =35 \,\rm cm \,s^{-1}$. It is consistent with the dust envelope
models.   The cone has an
additional feature in the form of an overdensity in velocity space
from 14 to $22\,\rm cm s^{-1}$, producing an elliptical band that, in
projection, resembles Arc 1.  The bottom edge of the cone, made
brighter by projection, resembles Arc 2.  The cone opening angle
increases with velocity from $33^\circ$ to $40^\circ$ to produce the
curvature observed in the second feature; this will be discussed
below.

We emphasize that this is not a best--fit model, but merely a 
representation of the outburst that agrees with our previous
fits of the dust envelope and inferences about the arcs.   A notable
oversimplification is that we assume the ejecta to be symmetric around
the central axis, which is certainly unjustified for a glancing
impact.  Such non--uniformity may explain why only parts of the ring
feature are visible, and why the brightest part is not the part viewed
in strongest projection. We also assume no gravity; 
when gravity corresponding to a massive \Atwo\ variant, modeled as a
sphere $100\,\rm m$ in radius with density $3500\,\rm kg\,m^{-3}$ is
introduced, the entire debris ensemble is pulled to the left by about
an arcsecond, arguing against such a massive body.  Less massive variants
of the nucleus, like those considered in Paper I, do not exhibit this shift.
Another shortcoming of this simple model is that the arc features
touch but do not fully cross, unlike the actual data. This can be
ameliorated by making the opening angle of the debris ring
larger than that of the cone, but there is no obvious physical reason
to do this.

A final limitation of this simple model is the fact that there is a
consensus that impact debris cones tend to widen at lower velocities
\citep{Richardson.Icarus.2007.ejecta}, whereas our cone is made
narrower to make it agree with the observed curvature of the debris
feature.  However, some laboratory studies
\citep{Anderson.JoGR.2003.ejecta,Cintala.1999.ejecta} find that the
cone re--narrows at the final low--velocity stage of excavation, which
is the regime relevant to the visible debris.

When we continue the integration to the 2010 March, April, and May
dates of the later HST images of \cite{Jew+10}, then
the development of our model subjectively matches that of \Atwo, 
with the features becoming narrower and trailing further behind
the nucleus.

Despite its {\it ad hoc} nature, this model shows that much of the
observed structure can be roughly replicated using only the best-fit
cone of the overall dust envelope, with the addition of non-uniformity
in velocity and a velocity variation in opening angle. Unfortunately,
a full fit of a cone plus ring model would be extremely difficult: it
would require optimizing over at least six parameters, would not
account for asymmetries ({\it e.g.} local terrain), would be
contaminated with $\beta>0$ dust, and would not have any obviously
correct merit function for the fit.

% this simimg has
%   vel-min=14e-5 vel-max=22e-5  theta=74 phi=-20 angle=40
%   thetab=74 phib=-20 angleb=20, no gravity
%
%  

\subsection{Physical implications of ejecta geometry}

Cratering phenomena and ejecta distributions have been studied
extensively using scaling laws based on dimensional arguments,
hydrodynamic simulations, and laboratory experiments (e.g.~R2007, and
\cite{HolsappleImactScalingReview1993}).  Our finding of
a $\alpha_c\approx45^\circ$ hollow cone is in broad agreement with a
number of sources cited in R2007, which find ejection angles
$\psi=90^\circ-\alpha_c$ ranging from $35^\circ$ to $63^\circ$,
corresponding to $\alpha_c= 55^\circ$ to $27^\circ$.  In laboratory
experiments involving the impact of high--speed projectiles into sand,
the most common value appears to be $\psi\approx 50^\circ$, in
agreement with our value $\alpha_c\approx40^\circ$
\citep{Anderson.JoGR.2003.ejecta,Cintala.1999.ejecta}.

Henceforth, we will use the R2007 formulation of cratering to discuss
some interpretations of our modeling of the observed phenomena.  We
strongly emphasize, however, that many aspects of cratering,
particularly late--stage effects, are poorly understood, and that our
conclusions hinge on the uncertain validity of these 
models. The standard cratering formulation of R2007 is almost
certainly an oversimplification that is at odds with experiment.
Nevertheless, we believe that the arguments are at least qualitatively
valid, and provide physical motivation for aspects of the simple
model in the previous section.

\cite{Jew+10} assume a mm to cm range of particle sizes and find that
the visible ejecta correspond to a sphere having a volume $ V_{\rm ej}
\in[2\times 10^4 , 2\times 10^5]\,\rm m^3$, so that the radius $R_c$
of the crater is $R_c\approx {V_e}^{1/3}\sim 30$ to $\sim 60\,\rm m$
from Equation 11 of R2007.  Similarly, in Paper I, we measure
particles spanning diameters of 1 to 20 mm, and a slightly larger
volume of ejecta of $2.7\times10^5\,{\rm m^3}$,  also giving a crater
with $R_c\sim 60\,{\rm m}$.

The accuracy of the volume estimate varies linearly with the accuracy
of the estimated debris particle size, so the accuracy of the crater
radius $R_c$ is much better, going as the $\frac{1}{3}$ power of the
particle size, which will be useful below.

\begin{figure*}[htb]
\includegraphics[width=9cm]{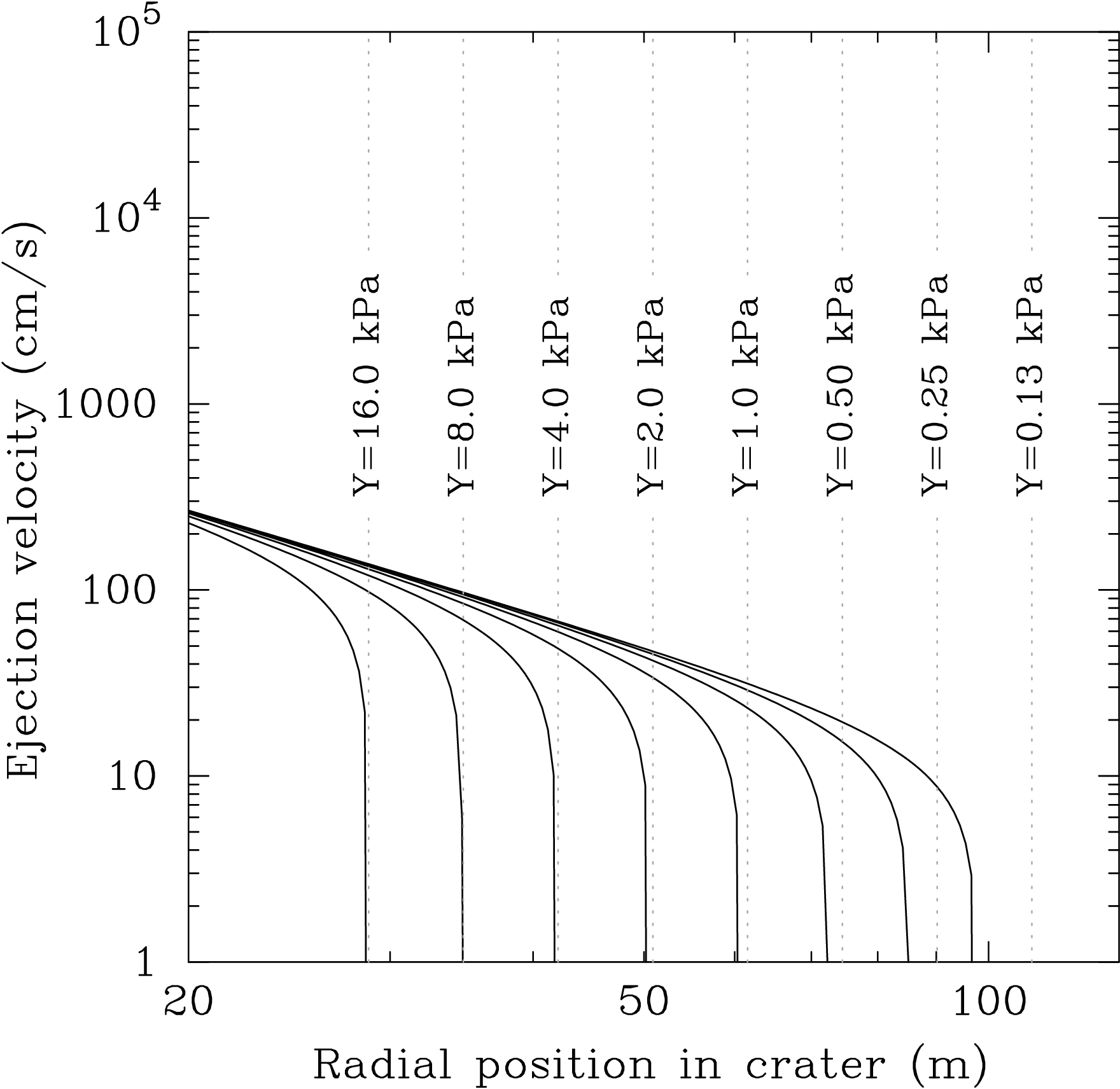} \hspace {1cm}
\includegraphics[width=9cm]{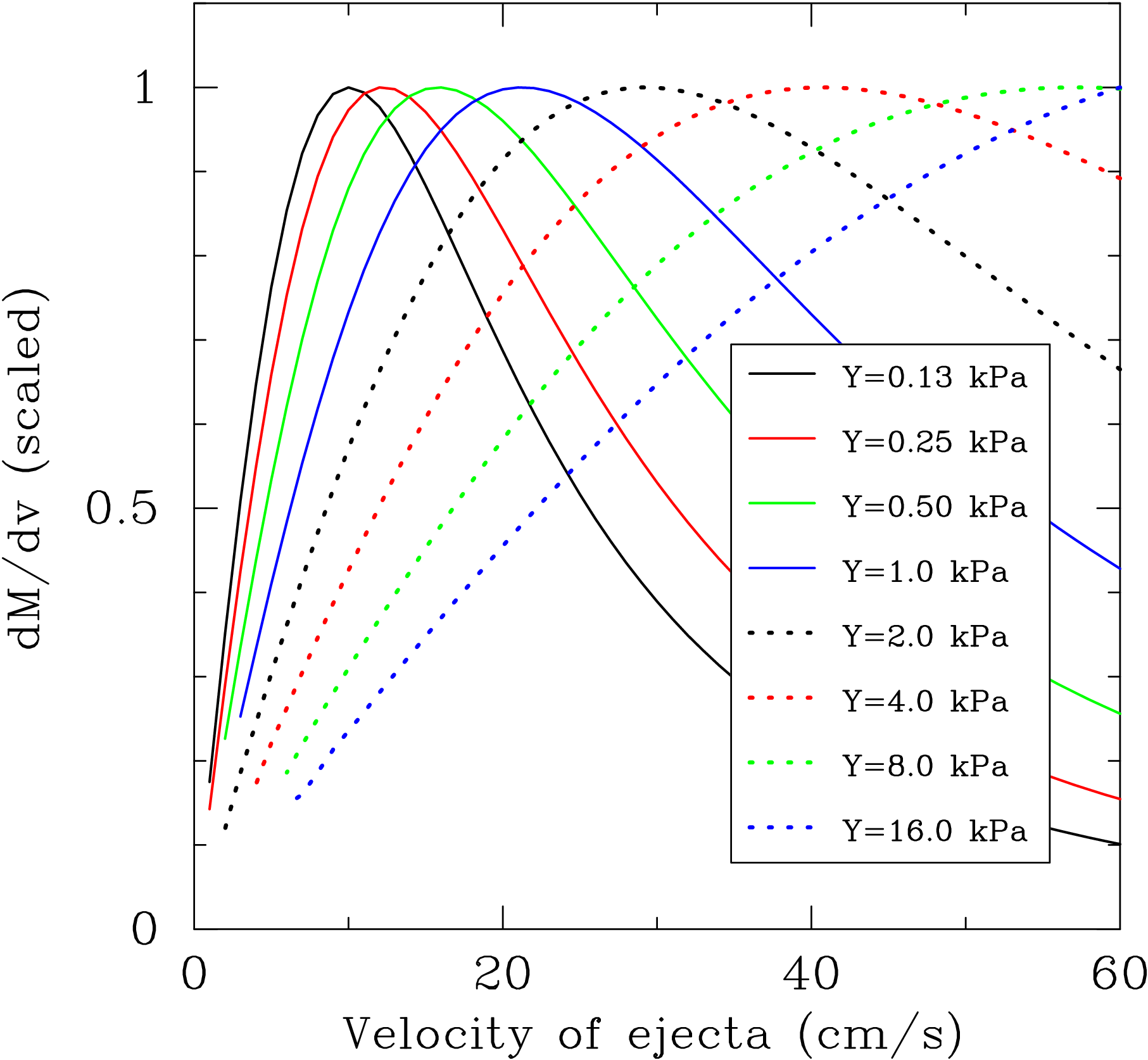} 
\caption{\label{fig:cratering} Left: Right: the ejection velocity (not
  corrected for $v_e$) as a function of the radial position in
  the crater, with a terminus at the strength--limited crater radius
  (dotted lines at each strength $Y$).  Each solid curve corresponds
  to one value of $Y$.  We used an $85\,\rm m$ target radius, an
  $3000\,\rm kg\,m^{-3}$ target density, and a $0.7\,\rm m$ impactor
  moving at $3\,\rm km\,s^{-1}$, close to the nominal values from
  Paper I.  Note that for small $Y$ crater growth stops before the
  corresponding dotted strength limit as gravity becomes relatively
  important at low strengths. Right: the differential distribution of
  ejected mass as a function of ejection velocity (corrected for
  $v_e$) for various target material strengths $Y$.  Only the
  lower target strengths correspond to the low debris velocities
  observed.}
\end{figure*}

In the simplified model of cratering, the crater radius $r_c$ grows in
a outflow of material, and debris is emitted at a velocity $v_e\propto
{r_c}^{-\frac{1}{\mu}}$, with $\mu\in[0.41,0.55]$, until a
strength--limited or gravity--limited crater volume and radius are
reached (Equations 13 and 37 in R2007).  The first panel of
Fig.~\ref{fig:cratering} illustrates this for a model of \Atwo\
resembling our estimate in Paper I: the radius of the target is
$85\,\rm m$, the density is $3000\,{\rm kg\,m^{-3}}$, and the
$r_i=0.6\,\rm m$ impactor, having the same density, strikes at
$v_i=3\,{\rm km\,s^{-1}}$.  The impactor is made slightly smaller and
slower than our canonical values of $r_i=1\,\rm m$ and $v_i=5\,{\rm
  km\,s^{-1}}$ to give crater sizes on the order of the $60\,\rm m$
estimate, rather than being larger than \Atwo.  In this figure, the
ejection velocity (ordinate) falls with crater radius (abscissa),
until the velocity plummets to zero at some critical strength-limited
radius (dotted vertical lines) given by the strength $Y$, where the
flow is no longer sufficiently energetic to break the material.  At
the lowest strengths, the crater radius never reaches
strength--limited radius because gravity contributes to the truncation
of growth.  In both the strength--limited and gravity--limited
regimes, the velocity curve has a sharp break or knee at the truncation
radius.  In both regimes, the maximum amount of debris is ejected just
before the crater's limiting radius, by considering the area of the
crater at the moment of truncation.

The $v_e(r)$ curve plotted in the left panel of
Fig.~\ref{fig:cratering} may be numerically inverted to give $r(v_e)$,
and the amount of mass ejected at a particular velocity may then be
written as $dM/dv_e \propto dM/dr \, \, dv_e/dr \propto r^2(v_e)\, dv_e/dr$.  The sharp bend
in $v_e(r)$ gives a large value for the derivative $dv_e/dr$, which,
combined with the large value of ${r_e}^2$, produce a peak in debris
at low velocities.  This peak is shown in the right panel of
Fig.~\ref{fig:cratering}:  $Y\sim 0.5\, \rm kPa$  produces peaks
in the debris around the value of  $14\,\rm cm\,s^{-1}$ that we
attribute to Arc 1, and also gives a crater of an appropriate size.
The system is generally 
  
This low strength corresponds to a loose powdery surface, comparable
to lunar soil ($Y\sim 1\,\rm kPa$, \cite{MitchellLunarRegolith1972})
and snow ($Y\sim 0.5\,\rm kPa$, \cite{SommerfieldSnowTensile1974}).  A
caveat is the fact that the concept of material strength in impacts is
not well defined, so the $Y$ given is a formal value with an uncertain
physical interpretation.  This entire approach should be viewed with
caution, because the cratering physics is greatly oversimplified, and
because the parameters of the system, though chosen from within the
the range indicated by observations, were tuned to give the desired
velocity peak and crater size.  Nevertheless, Fig.~\ref{fig:cratering}
suggests that high material strengths will not produce a peak of
material at a low velocity, but low strengths can be made {\it
  qualitatively} consistent with the physical picture that emerges
from orbit modeling.

Throughout, we have assumed a symmetric ejection event, corresponding
to a vertical or near---vertical impact.  In an oblique impact that is
more than $60^\circ$ from the vertical, debris becomes preferentially
distributed in the downrange direction, and a gap in the cone appears
in the direction from which the impactor arrived \citep[][and
R2007]{PierazzoObliqueImpactRev2000}.  The features observed in
Fig.~\ref{fig:crossOrbits} could be one side of an asymmetrical impact
cone, implying that the impactor arrived from $\theta_{\rm A2}\sim
0^\circ$, outside \Atwo's orbit.  Because the features are close to the
directions of maximum projection (Fig.~\ref{fig:orbitProjection}), it
is hard to distinguish between an oblique impact and a symmetric cone
that is selectively enhanced by the viewing geometry, as in
Fig.~\ref{fig:toyconefeatures}.  Equations 44, 45, and 46 of R2007
provide an approximation to the modifications of the ejection angle
and velocity resulting from an oblique impact, but these variations
turn out to be much too small to be visible, and cannot account for
the curvature of the second cross feature (Fig.~\ref{fig:crossOrbits},
right).

In conclusion, it is very plausible that the ring--shaped feature
(peak in the velocity distribution) suggested by the first panel of
Fig.~\ref{fig:orbitProjection} and by Fig.~\ref{fig:toyconefeatures}
is in fact related to the expected peak in the debris velocity
distribution: it may be the last, slowest, and most abundant debris
before the cratering process halted.  In this respect, it differs from
the Tempel 1 Deep Impact result
\citep[e.g.]{HolsappleDeepImpactCratering2007}, where the
tensile-strength limited velocity was apparently below the escape
speed, and the plume never detached. A further argument for plume
detachment in the \Atwo\ data is the visible gap separating the
nucleus and debris trail (HST image, Fig.~\ref{fig:toyconefeatures})

\section{Summary and discussion}

\newtext{ The peculiar collection of debris surrounding \Atwo\ has several
  possible explanations.  As originally pointed out by
  \citet{Jew+10}, it may arise from disruption after rotational
  spin--up, from prolonged sublimation activity, or from a collision,
  the hypothesis examined in this paper.  We found broad agreement
  with a conical ejection event, and the large particles argue against
  sublimation, but we did not specifically find disagreement with a
  spin--driven disruption.  To confirm or rule out this last
  possibility, similar orbit modeling could be applied to the ejecta
  distributions produced by spin--disruption.  Naively, a $200\,\rm m$ body spinning at the
  shortest 2.1 hour period permitted for a non--cohesive rubble pile would have a
  surface speed of $8 \,{\rm cm\,s^{-1}}$, far less than the observed
debris velocity of  $\sim30 \,{\rm cm\,s^{-1}}$.   However, a faster
cohesive body cannot be ruled out \citep{2007Icar..187..500H}}.

In this paper, we examined whether the January 2010 trail behind \Atwo could be
explained with a conical ejecta distribution from an impact event.
Specifically:

\begin{enumerate}

\item We performed fit (\S\ref{sec:dust}) using binary filling of a
  drawn-by-eye dust envelope, finding that two islands in the space of
  ejection direction yield $\alpha_c\sim40^\circ$ half angle ejecta
  cones that fill the observed dust envelope.  A $\sim 40^\circ$ cone is in
  agreement with cratering theory and experiment.  The two islands
  correspond to rotating the cone by one full width, so their $\sim
  90^\circ$ separation is further evidence of the validity of the d
  $\alpha_c\sim40^\circ$ cone solution -- {\it i.e} there is something
  in the system with a characteristic angular scale of $\sim
  90^\circ$.  The main narrow bright trail extending directly from the
  nucleus corresponds to a direction on the cone that is more aligned
  with the effect of solar pressure, resulting in coherent motion
  rather than solar pressure driven spreading.

\item We created a second set of models (\S\ref{sec:dustcomplex}),
  using power law distributions for the dust size, velocity, and
  velocity--cutoff, and employing an absolute value deviation metric
  between the image and the model.  We found that the same two
  ejection geometries provided the best fit, suggesting that this
  result is robust against the assumptions of the fitting method.
  These models gave a dust size exponent that matched previous
  results, but gave an excess of dust at large velocities relative to
  cratering theory.  They also placed higher velocity dust at the
  locations of the the N and SE diffuse features
  (Fig.~\ref{fig:schematic}).

\item We argue that the two bright arc features
  (Fig.~\ref{fig:schematic}) must consist of sheet--like or line--like
  ejection (\S\ref{sec:arcs}).  We integrated the trajectory of large
  ($\beta=0$) dust particles in all possible directions, to see which
  directions could plausibly land on the two bright arc features.  We
  found that the orbits corresponding to the bright arcs are low
  velocity debris lying on the same $\alpha_c\approx40^\circ$ cones
  found in the dust fits (Fig.~\ref{fig:crossOrbits}), which did not
  use knowledge of the features.  This agreement provides independent
  support for a $\alpha_c\approx40^\circ$ ejection cone.  We argue
  that one of the features is an edge--brightened region of cone
  enhanced by an alignment of the line of sight with the local
  velocity vector, and the other may be a concentration of debris in
  velocity.  The features thus result from debris coincident with the 
  cone, viewed in projection.  We construct a simple model
  using these ideas that {\it qualitatively} reproduces some of the
  salient features of \Atwo.

\item We apply standard cratering theory to argue that a peak
  quantity of debris at a low non--zero velocity is a natural
  consequence of an impact.  If the second arc feature corresponds to
  such a peak in the velocity distribution, it is agreement with an
  impact into loose regolith.

\end{enumerate}

We argue that from several different perspectives, a consistent view of \Atwo's
activity emerges: it is probably the result of a single impact into a
loose surface, throwing debris outward in a well--known hollow
$\sim40^\circ$ half-opening angle conical pattern, in qualitative
agreement with theoretical and laboratory studies of impact cratering.
Further work will require addressing asymmetries in the impact, and
more complicated velocity distributions.  Such improvements may be
constrained by the limited available data, which are contaminated by artifacts,
and by the unknown effects of the target's terrain and topography.

\begin{acknowledgements}
This material is based upon work supported by the National Aeronautics
and Space Administration through the NASA Astrobiology Institute under
Cooperative Agreement No. NNA09DA77A issued through the Office of
Space Science.  We would like to thank the referee for several helpful suggestions.
\end{acknowledgements}

% for the bibliography, at the end
\bibliographystyle{aa} 
\bibliography{P2010A2II}

\end{document}